\documentclass[aps,prb,preprint,superscriptaddress,showpacs]{revtex4-1}

\usepackage{graphicx}% Include figure files
\usepackage{dcolumn}% Align table columns on decimal point
\usepackage{amsthm}
\usepackage{amsmath}
\usepackage{bm}% bold math
\usepackage{longtable}
\usepackage{color}
 
\begin{document}

\title{Thickness dependence of electron-electron interactions in topological $p$-$n$ junctions}

\author{Dirk Backes}
\email{d.backes@lboro.ac.uk}
\affiliation{Cavendish Laboratory, University of Cambridge, J. J. Thomson Avenue, Cambridge CB3 0HE, United Kingdom}
\affiliation{Department of Physics, Loughborough University, Epinal Way, Loughborough LE11 3TU, United Kingdom}
\author{Danhong Huang}
\affiliation{Air Force Research Laboratory, Space Vehicles Directorate, Kirtland Air Force Base, New Mexico 87117, USA}
\author{Rhodri Mansell}
\affiliation{Cavendish Laboratory, University of Cambridge, J. J. Thomson Avenue, Cambridge CB3 0HE, United Kingdom}
\author{Martin Lanius}
\affiliation{Peter Gr\"unberg Institute (PGI-9), Forschungszentrum J\"ulich, 52425 J\"ulich, Germany}
\author{J\"orn Kampmeier}
\affiliation{Peter Gr\"unberg Institute (PGI-9), Forschungszentrum J\"ulich, 52425 J\"ulich, Germany}
\author{David Ritchie}
\affiliation{Cavendish Laboratory, University of Cambridge, J. J. Thomson Avenue, Cambridge CB3 0HE, United Kingdom}
\author{Gregor Mussler}
\affiliation{Peter Gr\"unberg Institute (PGI-9), Forschungszentrum J\"ulich, 52425 J\"ulich, Germany}
\author{Godfrey Gumbs}
\affiliation{Department of Physics and Astronomy, Hunter College of the City University of New York, 695 Park Avenue, New York, New York 10065, USA}
\author{Detlev Gr\"utzmacher}
\affiliation{Peter Gr\"unberg Institute (PGI-9), Forschungszentrum J\"ulich, 52425 J\"ulich, Germany}
\author{Vijay Narayan}
\email{vn237@cam.ac.uk}
\affiliation{Cavendish Laboratory, University of Cambridge, J. J. Thomson Avenue, Cambridge CB3 0HE, United Kingdom}

\date{\today}

\begin{abstract}

Electron-electron interactions in topological $p$-$n$ junctions consisting of vertically stacked topological insulators are investigated. $n$-type $\mathrm{Bi_2Te_3}$ and $p$-type $\mathrm{Sb_2Te_3}$ of varying relative thicknesses are deposited using molecular beam epitaxy and their electronic properties measured using low-temperature transport. The screening factor is observed to decrease with increasing sample thickness, a finding which is corroborated by semi-classical Boltzmann theory. The number of two-dimensional states determined from electron-electron interactions is larger compared to the number obtained from weak-antilocalization, in line with earlier experiments using single layers. 
\end{abstract}

\pacs{73.20.-r, 73.25.+i, 73.50.-h}

\maketitle

%------------------------------------------------------------------------------------------------------
\section{Introduction}

Topological insulators are fascinating materials with conducting surfaces, harboring electronic states with a Dirac-like bandstructure\cite{Hasan:2010}. Large spin-orbit interaction together with time reversal symmetry cause the topological nature of these surface states (TSS), manifesting itself in the suppression of backscattering and leading to the weak-antilocalization effect (WAL) and to spin-momentum coupling. Furthermore, magnetic topological insulators exhibit the quantum anomalous Hall (QAH) effect\cite{Haldane:1988, Yu:2010, Chang:2013}, characterized by dissipationless chiral currents. These properties of topological insulators have attracted great attention because of their potential applications in energy-efficient electronics and quantum computing.

The analysis of the topological properties is complicated by the non-zero conductivity of the bulk\cite{Analytis:2010, Checkelsky:2011, Taskin:2012}, which often dominates the overall transport characteristics. Several methods have been devised to suppress the bulk contribution, such as doping~\cite{Chen:2009, Kong:2011, Zhang:2011, Weyrich:2015}, gating~\cite{Chen:2010, Chen:2011, Checkelsky:2011, Steinberg:2011}, and reducing the thickness of the layer~\cite{Jiang:2012}. A relatively unexplored but elegant method is to combine an electron and hole dominated material to form a $p$-$n$ junction, and thus creating a depletion layer at the interface~\cite{Zhang:2013, Eschbach:2015, Backes:2017}.

The $\pi$-Berry phase of the Dirac fermions gives rise to quantum corrections of the conductivity, with a  magnetic field and temperature dependence resembling the WAL effect. By analyzing of the WAL in topological $p$-$n$ junctions the transport through TSS and bulk states was disentangled~\cite{Backes:2017}. Additional modifications of the conductivity are caused by electron-electron interactions (EEI), originating from an effective decrease of the electron density at the Fermi level \cite{Lee:1985,Altshuler:1998, Koenig:2013, Lu:2014}. The combined study of both WAL and EEI can reveal information about spin (EEI) and orbital (WAL) part of the electron wave function to transport~\cite{Checkelsky:2009}.

%Jiang et al. \cite{Jiang:2012} explain a minimum in peak-width distribution of Landau levels with EEI, limiting the quasi-particle lifetime. Takagaki et al. \cite{Takagaki:2012} see one 2D-channel in Cu-doped BiSe when they analyse WAL, but two from EEI. In contrast to that, the same group obtains two 2D-channels in SbTe-films\,\cite{Takagaki:2012b}, both using WAL and EEI. 

\begin{figure}
	\centering
	\includegraphics[width=\columnwidth]{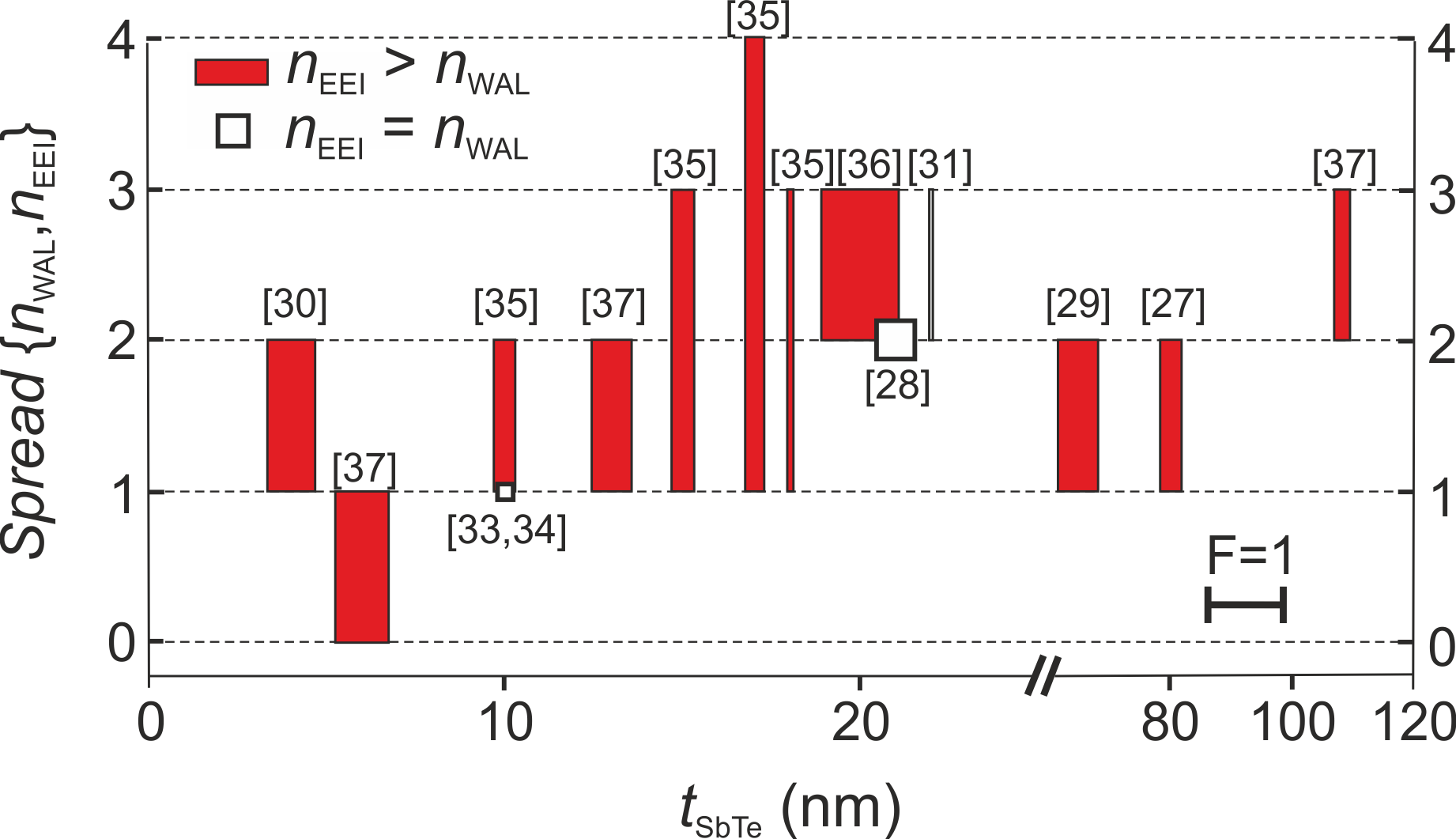}
	\caption{Comparison of the number of 2D channels from WAL ($n_{\mathrm{WAL}}$) and EEI ($n_\mathrm{EEI}$) as a function of the layer thickness. The values are taken from literature, with the references given in brackets. The bars indicate the spread between $n_\mathrm{EEI}$ (top) and $n_{\mathrm{WAL}}$ (bottom). Squares indicate experiments where $n_{\mathrm{EEI}}=n_{\mathrm{WAL}}$. The widths of the bars are proportional to the screening factor $F$ (see scale bar in the bottom right). }
	\label{fig1}
\end{figure}

Especially the number of 2D states $n$ is of utmost interest, since it can provide evidence of the topological nature of a TI \cite{Narayan:2016,Nguyen:2016}. By careful observation of either the WAL or EEI, a value for $n$ can be gained\cite{Wang:2011, Takagaki:2012, Takagaki:2012b, Chiu:2013, Roy:2013, Takagaki:2014, Dey:2014, Jing:2016, Trivedi:2016, Kuntsevich:2016, Sahu:2018,Wang:2016}. It turns out that in single layer TI, $n_\mathrm{EEI}$  tends to be larger than $n_\mathrm{WAL}$ \cite{Wang:2011, Takagaki:2012, Chiu:2013, Roy:2013, Takagaki:2014, Dey:2014, Kuntsevich:2016, Sahu:2018,Wang:2016} (see Fig. \ref{fig1} and Tab.\,\ref{tab1}). It seems that surface states on the top and bottom contribute independently to EEI but that, under certain circumstances, they appear to be coupled when the WAL effect is concerned. The physical origin of this coupling effect remains elusive. Predominantly in very thin layers only one 2D state contributes to WAL\cite{Roy:2013, Jing:2016, Trivedi:2016, Wang:2016, Kuntsevich:2016}. Thicker films tend to be decoupled when WAL is concerned and therefore exhibit a higher number of 2D-channels\cite{Takagaki:2012b, Wang:2016, Takagaki:2014, Sahu:2018}. Microflakes \cite{Chiu:2013} and hot wall epitaxy deposited layers\cite{Takagaki:2012} are exceptions where coupling effects can be observed even at thicknesses $>$\,60\,nm. A combined study of the WAL and EEI in TI-multilayers is entirely missing.

%In EEI experiments, on the other hand, the different transport channels act independently and their number is almost always higher compared to WAL. The strength of the EEI effect depends on the screening of the Coulomb interaction, expressed by the screening factor $F$\cite{Lee:1985}. $F$ should in principle take on values between 0 and 1, with lower values corresponding to lower screening, leading to stronger EEI effects. A negative screening factor is not allowed and can be interpreted as an indication for more than one channel. Nevertheless, it also has been attributed to high spin-orbit coupling\cite{Chiu:2013}. 

In the following, we present the first investigation of the interplay of WAL and EEI in topological $p$-$n$ junctions. Conductivity corrections are measured at temperatures $<$ 10\,K as a function of temperature, magnetic field and sample thickness. The conductivity correction are used to find the number of 2D channels contributing to either EEI or WAL. Finally,  a semiclassical Boltzmann theory is derived to understand the thickness dependence of the conductivity corrections due to EEI. 
%------------------------------------------------------------------------------------------------------
\section{Experiment}

The $\mathrm{Bi_2Te_3/Sb_2Te_3}$-bilayers (BST) were grown using molecular beam epitaxy (MBE). Details of the MBE sample preparation can be found in Ref.~\onlinecite{Eschbach:2015}. The bottom $\mathrm{Bi_2Te_3}$-layer was $t_\mathrm{BiTe}$~=~6\,nm and the top $\mathrm{Sb_2Te_3}$-layers was 6.6\,nm (BST6), 7.5\,nm (BST7), 15\,nm (BST15), and 25\,nm (BST25) thick, respectively. The films were patterned into Hall bars which were $\mathrm{200\,\mu m}$ wide and $\mathrm{1000\,\mu m}$ long. Transport in these samples was measured in a He-3 cryostat at temperature down to 300\,mK while a perpendicular magnetic field could be applied using a superconductive magnet.

\begingroup
\begin{table}
\label{tab1}
\begin{tabular}{|c|c|c|c|}
\hline
 Ref.	&	Sample	&	Method	&	t/nm		\\
\hline	
Roy et al. \cite{Roy:2013}				&	BiTe			&	MBE	&	4\\
Wang et al. \cite{Wang:2016}			&	BiSe			&	SP	&	6-108\\
Jing et al. \cite{Jing:2016}				&	BiSe			&	MBE	&	10\\
Trivedi et al. \cite{Trivedi:2016}			&	BiTeS 		&	Flakes		&	10\\
Kuntsevich et al. \cite{Kuntsevich:2016}	&	BiSe films		&	MBE	&	10-18\\
Sahu et al. \cite{Sahu:2018}			&	BiSe films		&	SP	&	20\\
Takagaki et al. \cite{Takagaki:2012b}	&	SbTe films	&	MBE	&	21\\
Takagaki et al. \cite{Takagaki:2014}		&	SbTe			&	MBE	&	22\\
Chiu et al. \cite{Chiu:2013}			&	BiTe 			&	Flakes	&	65\\
Takagaki et al. \cite{Takagaki:2012}		&	Cu-doped BiSe	&	HWE	&	80\\

\hline
\end{tabular}
\caption{Sample details of experiments reporting both on WAL and EEI. Most results are reported on thin films grown by molecular beam epitaxy (MBE) and sputtering (sp) and a few by hot wall epitaxy (HTW) and on microflakes.}
\end{table}
\endgroup

%------------------------------------------------------------------------------------------------------
\section{Results}

\begin{figure}
	\centering
	\includegraphics[width=\columnwidth]{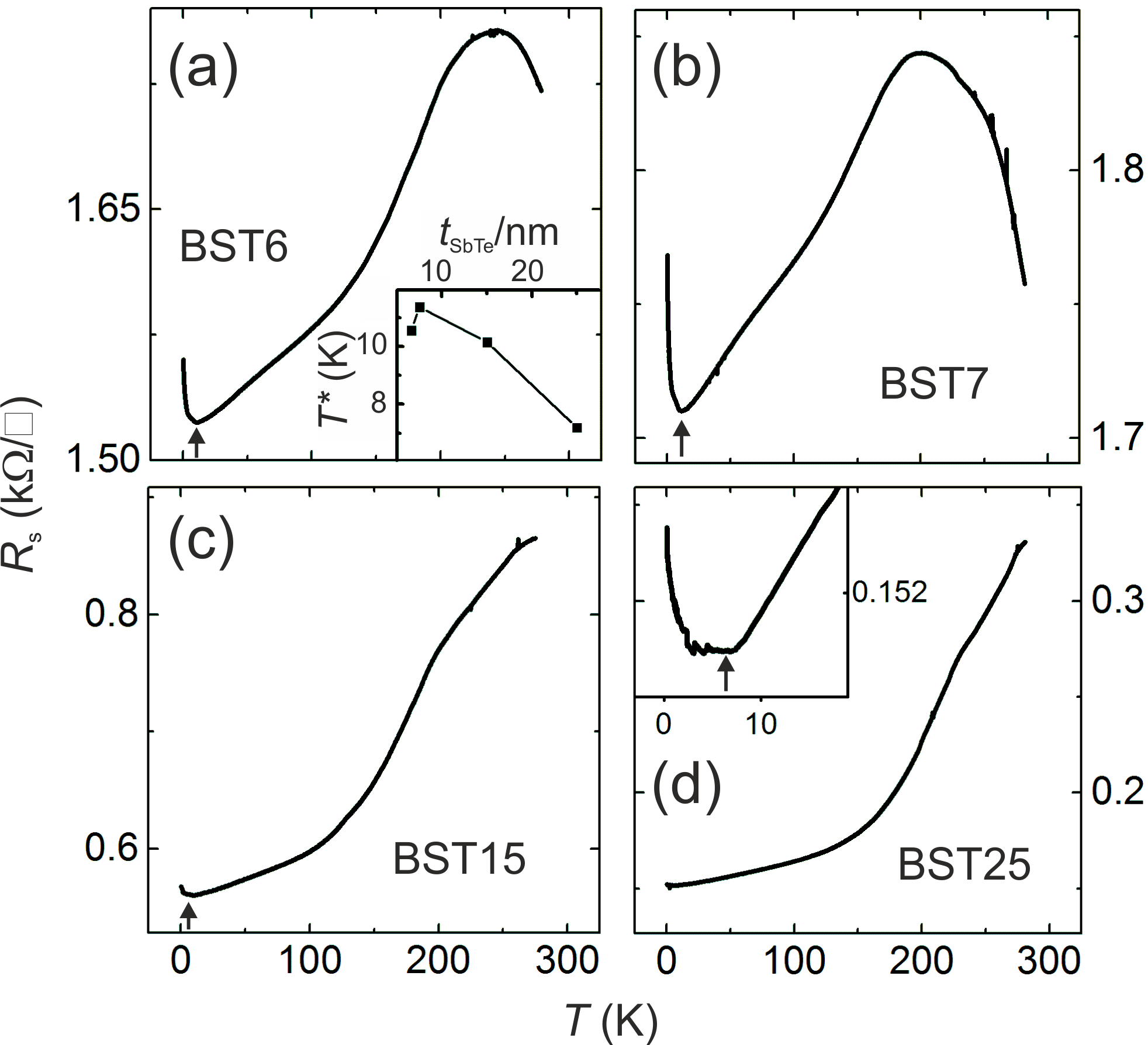}
	\caption{(a)-(d) Sheet resistance $R_\mathrm{s}$ dependance on temperature for four different samples. The arrows indicate the transition temperature $T^\mathrm{*}$. Insert in (a) Transition temperature $T^\mathrm{*}$ dependence on thickness of the $\mathrm{Sb_2Te_3}$-layer}
	\label{fig2}	
\end{figure}

In Fig. \ref{fig2} the sheet resistance $R_\mathrm{s}$ during cooldown is shown for all sample thicknesses. Metallic behavior is dominant, except for the thinnest samples, BST6 and 7, which are insulating between room temperature and 200\,K, where they become metallic. At base temperature (300\,mK) all samples are insulating, with the transition temperature between the metallic and insulating phase, $T^\mathrm{*}$,  found to be between 7 to 11\,K, depending on the sample thickness (see insert in Fig.\,\ref{fig2}(a)). 

The temperature range below $T^\mathrm{*}$ is explored in more detail in Fig.\,\ref{fig3} for each sample thickness. The temperature was increased in small steps starting at base temperature of 300\,mK, taking care for the temperature to stabilize. An external magnetic field was swept between 0 and 0.5\,T at each temperature step. Both longitudinal and transverse resistance were recorded from which the conductivity could be calculated. Only one field loop needed to be taken since the noise level was low.  

%The relative resistance change, compared between the change in the insulating phase at low temperature and the metallic phase, is largest for BST 6 and BST 7 (19\% and 57\%) and smaller for the thicker samples (1\% and 4\%). 

%------------------------------------------------------------------------------------------------------
\section{Discussion}

\begin{figure}
	\centering
	\includegraphics[width= \columnwidth]{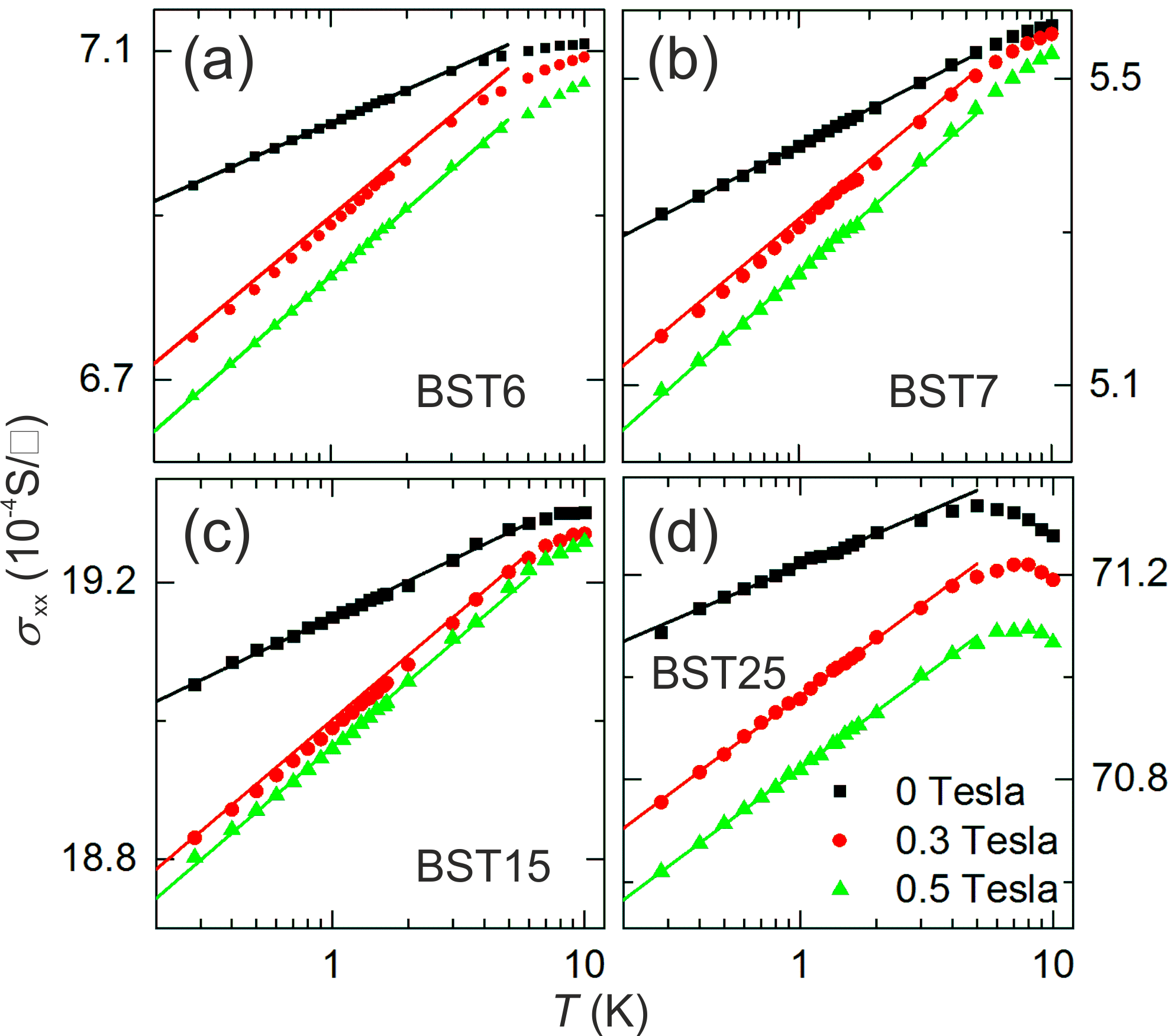}
	\caption{(a) - (d) Conductivity of four different samples at low temperature for three different perpendicular magnetic fields. Using a logarithmic scale for the temperature, the linear regions are fitted using Eqn.~\ref{EEI} (straight lines). The magnetic field leads to a change of slope, from which the screening and number of 2D channels can be derived.}
	\label{fig3}
\end{figure}

EEI originate from pairing of electrons at the Fermi energy and lead to a decrease in the carrier density, which in turn leads to a reduction of the conductivity. As can be seen in Fig.\,\ref{fig3}, the correction to conductivity due to EEI sets in below a transition temperature and exhibits a well-defined temperature dependence, given by \cite{Lee:1985}
\begin{equation}
\label{EEI}
\delta \sigma(T) = -\frac{e}{\pi h}n \left(1-\frac{3}{4}F \right) \ln \left(\frac{T}{T^{\star}}\right)
\end{equation}
where n is the number of 2D channels, $F$ the screening factor, and $T^\mathrm{*}$ the transition temperature. By applying Eq.~\ref{EEI} to the measured conductivity in Fig.~\ref{fig3} using $T^\mathrm{*}$ (see insert in Fig.~\ref{fig2}(a)), we obtain $f=n(1-3/4*F)$ from the slope of the temperature dependence. 

\begin{figure}
	\centering
	\includegraphics[width=\columnwidth]{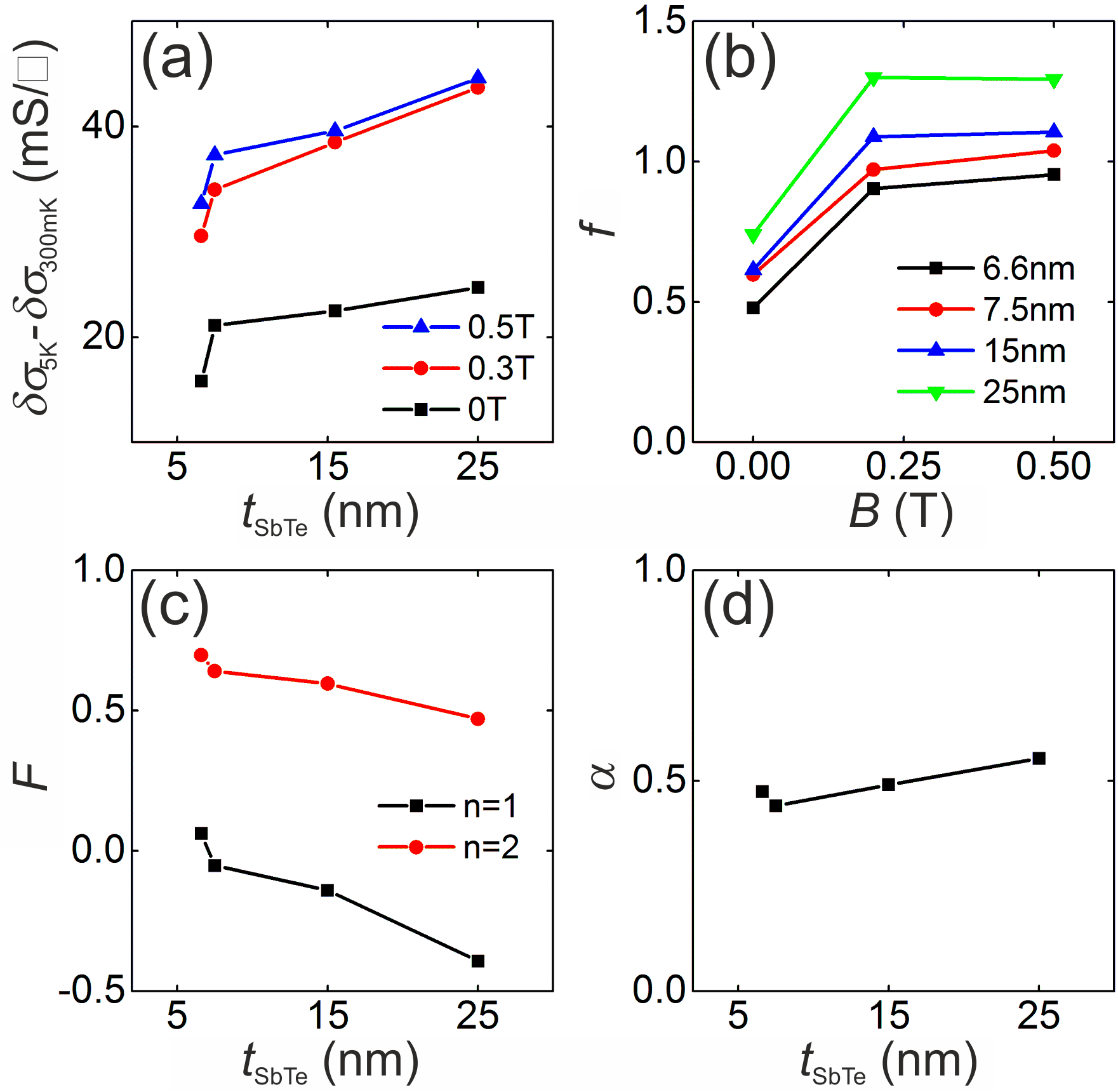} %6.5in
	\caption{(a) Difference of conductivity correction $\delta\sigma$ between 5\,K and base temperature as a function of the $\mathrm{Sb_2Te_3}$-thickness. (b) Change of the slope $f$ with an external, perpendicular magnetic field, as shown in Fig. \ref{fig3}. (c) The screening factor $F$ calculated from $f=n(1-3/4*F)$, asuming the number of 2D states $n$ is 1 (black squares) or 2 (red circles). The screening is negative for $n=1$ and between 0 and 1 for $n=2$, supporting the presence of more than one 2D channel. (d) Number of 2D channels $\alpha$ from WAL, obtained as described in the text. A value of 0.5 corresponds to one 2D channel.}
	\label{fig4}
\end{figure}

The overall change of the conductivity correction between base and transition temperature, $\delta\sigma_{\rm 5K}-\delta\sigma_{\rm 300mK}$, increases with sample thickness (see Fig.~\ref{fig4}(a)).

Fig.~\ref{fig4}(b) shows the change of $f$ when a magnetic field is applied perpendicular to the sample. The value of $f$ is smaller than 1 without magnetic field but rises to values close or above 1 at fields $\approx 0.2\,$T. This abrupt change reflects the disruption of phase coherence due to the magnetic field, impacting WAL. At fields $> 0.2\,$T, where WAL has disappeared~\cite{Backes:2017}, any change in conductivity can be attributed to EEI. $f$ saturates above this field (see Fig.~\ref{fig4}(b)) and is employed to investigate the underlying EEI it originates from. 
The screening parameter $F$ can be inferred from $f$ if $n$, the number of 2D states is known. $F$ can attain values between 0 (no screening) and 1 (strong screening). This condition cannot be fulfilled when $f$ is larger than 1 and $n=1$. Thus, to obtain an $F$ within the allowed range from our experimental results\cite{Takagaki:2012} we assume that $n >1$ (see Fig.~\ref{fig4}(c)).

For $n=2$ the screening factor $F$ decreases with thickness, from 0.73 for BST6 to 0.5 for BST25 (see Fig.~\ref{fig4}(c)). It cannot be excluded that $n>2$ but although the values of $F$ differ, the thickness dependence remains unchanged. This goes hand-in-hand with a similar thickness-dependent increase of the conductivity correction, since weaker screening means stronger EEI, hence larger $\delta\sigma$. In single layers, both a decrease~\cite{Kuntsevich:2016,Wang:2016} as well as an increase~\cite{Trivedi:2016} of $F$ with increasing thickness have been reported. The increase was attributed to a stronger screening due to the bulk states in thicker samples~\cite{Trivedi:2016}. 

To explain our results in light of these contradicting earlier observations, we derived a semi-classical Boltzmann theory for the topological $p$-$n$ junctions. The total conductivity (see Eqns.\,C18 and C19 in the Supplement\,\cite{SM} for its derivation) is given by
\begin{widetext}
\begin{eqnarray}
\tensor{\mbox{\boldmath${\sigma}$}}_{\rm tot}({\mbox{\boldmath${B}$}})=e\,\tensor{\mbox{\boldmath${\mu}$}}^\|_{\rm v}({\mbox{\boldmath${B}$}})N_{\rm A}A_{\rm h}\left[(L_{\rm A}-W_{\rm p})+\int^{W_{\rm p}}_{0} dz\,\exp\left(-\frac{\beta e\bar{\mu}_{\rm h}N_{\rm A}}{2\epsilon_{\rm 0}\epsilon_{\rm r}D_{\rm h}}\,z^2\right)\right]
-e\,\tensor{\mbox{\boldmath${\mu}$}}^\|_{\rm c}({\mbox{\boldmath${B}$}})N_{\rm D}A_{\rm e}\nonumber\\
\times\left[(L_{\rm D}-W_{\rm n})+\int^{W_{\rm n}}_{0} dz\,\exp\left(-\frac{\beta e\bar{\mu}_{\rm e}N_{\rm D}}{2\epsilon_0\epsilon_{\rm r}D_{\rm e}}\,z^2\right)\right]+e\,\tensor{\mbox{\boldmath${\mu}$}}^{\pm}_{\rm s}({\mbox{\boldmath${B}$}})\,\left(\frac{\alpha_0\Delta_{\rm 0}}{2\pi\hbar^2v_{\rm F}^2}\right)\left(L_{\rm A}-L_{\rm 0}\right)A_{\rm s}\,\nonumber\\
\label{eq2}
\end{eqnarray}
\end{widetext}
where $A_\mathrm{s}=\tau_\mathrm{s}/\tau_\mathrm{sp}$ and $A_\mathrm{e,h}=\tau_\mathrm{e,h}/\tau_\mathrm{p(e,h)}$. $\tau_\mathrm{s}$ and $\tau_\mathrm{e,h}$ are the energy relaxation and $\tau_\mathrm{sp}$ and $\tau_\mathrm{p(e,h)}$ the momentum relaxation time of the surface and bulk, respectively. $L_{\rm A,D}$, $N_{\rm A,D}$, $\bar{\mu}_{\rm h,e}$, $W_{\rm p,n,}$ and $D_{\rm h,e}$ are thickness, electron density, mobility, range of depletion zone and diffusion coefficient of the acceptor (donator) layer, respectively. $v_{\rm F}$ is the Fermi velocity of the surface states which are allowed to have a small band gap $\Delta_0$ due to hybridization at small thicknesses. $\alpha_0$ and $L_0$ are constants to be determined experimentally. The surface mobility is 

\begin{equation}
\tensor{\mbox{\boldmath${\mu}$}}_{\rm s}({\mbox{\boldmath${B}$}})=\frac{\mu_{\rm 1}}{1+\mu^2_{\rm 1}B^2}\,
\left[\begin{array}{cc}
1 & \mu_{\rm 1}B\\
-\mu_{\rm 1}B & 1\\
\end{array}\right]\ ,
\label{eq3}
\end{equation}
with $\mu_{\rm 1}=e\tau_{\rm sp}v_{\rm F}^2/\Delta_0=e\tau_{\rm sp}v_{\rm F}^2/2k_{\rm B}T_0$. For weak magnetic field, we have $\mu_\mathrm{1}B \ll 1$, $\mu_\mathrm{xx} = \mu_\mathrm{yy} = \mu_\mathrm{1}$ and $\mu_\mathrm{xy} = -\mu_\mathrm{yx} = \mu_\mathrm{21}B$. 
  
 When $B\to 0$ the conductance correction  (see Eq.\,C20 in the Supplement \cite{SM}) is given by

 \[ 
\delta\sigma(T_{\rm e},u_{\rm s})\equiv\sigma_{\rm tot}(T_{\rm e},u_{\rm s})-\sigma_{\rm tot}^{\rm (0)}(T_{\rm e},u_{\rm s})
\]
\begin{eqnarray}
&=-\mu_0^s\left(\frac{\alpha_0\Delta_0}{2\pi\hbar^2v_F^2}\right)(L_A-L_0)\left[\frac{\tau_0^s(T_e,u_s)}{\tau_0^s(T_e,u_s)+\tau^{s}_{\rm pair}(T_e,u_s)}\right]\nonumber\\
&\approx -\sigma_0^s\left[\frac{\tau_0^s(T_e,u_s)}{\tau^{s}_{\rm pair}(T_e,u_s)}\right]\ ,
\label{eq4}
\end{eqnarray}

where $\mu_{\rm 0}^{\rm s}=e\tau_{\rm 0}^{\rm s}v_{\rm F}^2/\Delta_0=e\tau_{\rm 0}^{\rm s}v_{\rm F}^2/2k_{\rm B}T^*$, $\sigma_{\rm 0}^{\rm s}$ and $\tau_{\rm 0}^{\rm s}$ 
are the mobility, conductivity and energy-relaxation time, respectively, of surface electrons in the absence of EEI. 

Here, $\tau^{\rm s}_{\rm pair}(T_{\rm e},u_{\rm s})$ is the additional electron-electron pair scattering contribution to the inverse energy relaxation time (see Eq.\,C16 and C17 in the Supplement\,\cite{SM}), given by 

\[
\frac{1}{\tau^{\rm s}_{\rm pair}(T_{\rm e},u_{\rm s})}=\frac{1}{n_{\rm 0}{\cal A}}\sum_{{\bf k}_\|}\,\frac{f^s_{{\bf k}_\|}}{\tau^s_{\rm pair}({\mbox{\boldmath$k$}}_\|)}
\approx\frac{1}{16\pi^4\hbar n_{\rm 0}}\left(\frac{e^2}{2\epsilon_0\epsilon_{\rm b}}\right)^2
\]
\[
\times\int\limits_{q_0}^{1/\delta_{\rm s}}
\frac{dq_\|}{q_\|}\,
\left\{1-\left(\frac{e^2q_\|}{2\epsilon_0\epsilon_{\rm b}}\right)\frac{32k_{\rm B}T^*}{\pi\hbar^2\Gamma_0^2}\left(\frac{T^*}{T_{\rm e}}\right)
 D \right\} \int d^2{\mbox{\boldmath$k$}}_\|\,f^{\rm s}_{{\bf k}_\|}
\]
\[
\times\int d^2{\mbox{\boldmath$k$}}'_\|
\left[f^{\rm s}_{{\bf k}'_\|}(1-f^{\rm s}_{{\bf k}_\|^{-}})(1-f^{\rm s}_{{\bf k}^{'+}_\|})
+f^{\rm s}_{{\bf k}^{-}_\|}f^{\rm s}_{{\bf k}_\|^{'+}}(1-f^{\rm s}_{{\bf k}'_\|})\right]
\]
\begin{equation}
\times\frac{\Gamma_0/\pi}
{(\varepsilon^{\rm s}_{{\bf k}_\|}+\varepsilon^{\rm s}_{{\bf k}'_\|}-\varepsilon^{\rm s}_{{\bf k}_\|^{-}}-\varepsilon^{\rm s}_{{\bf k}^{'+}_\|})^2+\Gamma_0^2}\ ,
\label{eq5}
\end{equation}

where 

\begin{equation}
f^{\rm s}_{{\bf k}_\|}\approx \frac{2\pi\hbar^2v_{\rm F}^2\,n_0}{(k_{\rm B}T_{\rm e})^2(1+\Delta_0/k_{\rm B}T_{\rm e})}\,\exp\left(-\frac{\varepsilon^{\rm s}_{{\bf k}_\|}-\Delta_0}{k_{\rm B}T_{\rm e}}\right)\ ,\nonumber
\label{eq6}
\end{equation}

and $n_0=(m_{\rm s}^*/2\pi\hbar^2)E^{\rm s}_{\rm F}=(\Delta_0/2\pi\hbar^2v_{\rm F}^2)E^{\rm s}_{\rm F}=(k_{\rm B}T^*/\pi\hbar^2v_{\rm F}^2)E^{\rm s}_{\rm F}\sim\alpha_0(L_{\rm A}-L_{\rm 0})$.
We use $\gamma=+1$ and $q_0=\Gamma_0/\hbar v_{\rm F}$ as a cutoff for $q_\|\to 0$.  ${\bf k}_\|^{\pm}$ stands for ${\bf k}_\|\pm{\bf q}_\|$ and $D=C_0+\ln\left(T_{\rm e}/T^*\right)-1/2\ln2\left(T_{\rm e}/T^*\right)^2$. Here, pair scattering of bulk electrons will lead to reduction of total conductivity. 

Important conclusions can be drawn from these theoretical results. Firstly, for a weak magnetic field $B$, the longitudinal conductivity becomes independent of $B$, although the Hall conductivity depends on $B$ (see Eqns.\,\ref{eq2} and \ref{eq3}). Furthermore, Eqn.\,\ref{eq5} for the energy relaxation time indicates that both pair scattering and screening effects from EEI do not depend on $B$. This is a strong argument in favor analyzing EEI by applying a weak magnetic field, in order to separate quantum corrections due to WAL from $\delta\sigma$ (see Eqn.\,\ref{EEI} and Fig.\,\ref{fig4}(b)).

Secondly, the experimentally found strong increase of EEI with the sample thickness (see Fig.\,\ref{fig4}(a)) can be directly derived from the theory. Eqn.\,\ref{eq4} gives the dominant EEI-induced change in surface longitudinal conductivity at low $B$ fields and reveals its thickness dependence. 
On the one hand, we know that $\delta\sigma\propto\sigma_0^s\sim (L_{\rm A}-L_{\rm 0})$. On the other hand, we find that the ratio $\tau_0^s/\tau^s_{\rm pair}\propto n_0\sim (L_{\rm A}-L_{\rm 0})$. Overall, $\delta\sigma\propto (L_{\rm A}-L_{\rm 0})^2$ which for $(L_{\rm A}-L_{\rm 0})/L_{\rm 0}\ll 1$ leads to $\delta\sigma\propto L_{\rm A}$.  This linear relationship describes our experimental findings remarkably well (see Fig.\,\ref{fig4}(a)). Finally, bulk electrons can also screen impurity scattering of surface electrons, but it becomes insignificant due to the large separation between the surface layer and the center of film. 
%as can be seen from Eqs.\,(\ref{new-24}) and (\ref{a-2}). We know that $1/\tau^s_{\rm pair}$ has its density dependence of both $\sim n_0$ and $\sim n_0^2$ while $1/\tau^s_{\rm 0}\sim n_0$. Thus $\tau^s_{\rm 0}/\tau^s_{\rm pair}$ has a constant and a term $\sim n_0$. Because of $\sigma_0^s \sim (L_A-L_0)$,  $\delta\sigma_{tot}$ has terms both $\propto n_0$ and $\propto n_0^2$ with $n_0 \sim (L_A-L_0)^{-1}$. Thus $\delta\sigma_{tot}\sim L_A^2$ for $(L_A-L_0)/L_0\gg 1$. As $(L_A-L_0)/L_0\ll 1$, on the other hand, we find $\delta\sigma_{tot}(T_e,u_s)\propto L_A$.

%Meanwhile, we also find the ratio $\tau_0^s/\tau^s_{\rm pair}\propto n_0\sim (L_A-L_0)^2$, as can be seen from Eqs.\,C16 and D2 in the Supplement\,\cite{SM}.
%Although the screening due to electron-electron interaction can weaken the impurity scattering and increases the mobility, the conductivity is not affected by the momentum-relaxation time $\tau_{sp}$ of surface electrons.
%Even for two-dimensional electron gases in a quantum well, where they acquire a static dielectric function\,\cite{r5} $\epsilon_s(q_\|)\equiv\epsilon(q_\|,\omega=0)=1+q_s/q_\|$ with a Thomas-Fermi screening length $1/q_s$, the screened impurity scattering can also increase their conductivity.
 
The fact that $n=2$ indicates that 2 independent 2D channels are involved and stands in contrast to the results of WAL measurements (see Ref.~\cite{Backes:2017} and Fig.~\ref{fig4}(d)). This discrepancy between WAL and EEI has been reported in Cu-doped BiSe single layers \cite{Takagaki:2012} and attributed to a 2D bulk state. For SbTe single layers \cite{Takagaki:2012b}, it was speculated that one coupled state of top and bottom TSS dominates WAL, but that they contribute independently to EEI. It is not clear how coupling could be mediated in our bilayer samples, since the depletion layer at the interface separates the SbTe and BiTe layer. Therefore, it is more likely that the 2D bulk plays a role in EEI processes in our samples.

Lastly, we determine the WAL contribution form the difference between the saturated and zero field amplitude $\Delta f$. We have shown already that EEI is independent of the magnetic field, and thus the change of the slope of the $\delta\sigma$ with and without applied field can be attributed to WAL alone. The number of 2D states can be calculated using $\Delta f =p\times \alpha$ with $p=1$, which characterizes the temperature dependence of the coherence length (see Ref.~\cite{Backes:2017}). We obtain $\alpha \approx$ 0.5, i.e. that only one TSS is present at all thicknesses \cite{Backes:2017,Jing:2016, Trivedi:2016} (see Fig.~\ref{fig4}(d)). Since a TSS on the top surface has been confirmed in ARPES experiments~\cite{Eschbach:2015}, we conclude that the TSS  at the bottom must be disrupted.

In summary, topological $p$-$n$ junctions exhibit a rich set of transport characteristics related to their topological surfaces states. At low temperature, WAL and EEI compete in reducing the conductivity. The fact that EEI are unaffected by an external magnetic field was taken advantage of to determine the number of 2D channels. While exactly one was found from WAL, at least two are contributing to EEI. The growing presence of bulk states does not lead to stronger screening. On the contrary, conductivity corrections due to EEI are getting stronger with increase thickness. This effect could be understood withing a semiclassical Boltzmann theory. 

\medskip

\begin{acknowledgments}
D.B., D.R. and V.N. acknowledge funding from the Leverhulme Trust, UK, D.B., R.M., D.R., and V.N. acknowledge funding from EPSRC (UK). D.H. would like to thank the support from the Air Force Office of Scientific Research (AFOSR). G.M., M.L., J.K. and D.G. acknowledge financial support from the DFG-funded priority programme SPP1666. 
\end{acknowledgments}

\newpage

\draft
\title{Supplementary Material for ``Thickness dependence of electron-electron interaction in topological $p$-$n$ junctions''}

\author{Dirk Backes}
\email{d.backes@lboro.ac.uk}
\affiliation{Cavendish Laboratory, University of Cambridge, J. J. Thomson Avenue, Cambridge CB3 0HE, United Kingdom}
\author{Danhong Huang}
\affiliation{Air Force Research Laboratory, Space Vehicles Directorate, Kirtland Air Force Base, New Mexico 87117, USA}
\author{Rhodri Mansell}
\affiliation{Cavendish Laboratory, University of Cambridge, J. J. Thomson Avenue, Cambridge CB3 0HE, United Kingdom}
\author{Martin Lanius}
\affiliation{Peter Gr\"unberg Institute (PGI-9), Forschungszentrum J\"ulich, 52425 J\"ulich, Germany}
\author{J\"orn Kampmeier}
\affiliation{Peter Gr\"unberg Institute (PGI-9), Forschungszentrum J\"ulich, 52425 J\"ulich, Germany}
\author{David Ritchie}
\affiliation{Cavendish Laboratory, University of Cambridge, J. J. Thomson Avenue, Cambridge CB3 0HE, United Kingdom}
\author{Gregor Mussler}
\affiliation{Peter Gr\"unberg Institute (PGI-9), Forschungszentrum J\"ulich, 52425 J\"ulich, Germany}
\author{Godfrey Gumbs}
\affiliation{Department of Physics and Astronomy, Hunter College of the City University of New York, 695 Park Avenue, New York, New York 10065, USA}
\author{Detlev Gr\"utzmacher}
\affiliation{Peter Gr\"unberg Institute (PGI-9), Forschungszentrum J\"ulich, 52425 J\"ulich, Germany}
\author{Vijay Narayan}
\email{vn237@cam.ac.uk}
\affiliation{Cavendish Laboratory, University of Cambridge, J. J. Thomson Avenue, Cambridge CB3 0HE, United Kingdom}

\date{\today}

\maketitle
\appendix

\section{Bulk Boltzmann Moment Equation}

For an $n$-doped semiconductor bulk material,
we will start with the standard semiclassical Boltzmann transport equation for electrons within a conduction band $\varepsilon_c({\mbox{\boldmath$k$}})$ of a bulk. For this case, the electron distribution function $f_c({\mbox{\boldmath$r$}},\,{\mbox{\boldmath$k$}};\,t)$ satisfies

\[
\frac{\partial f_c({\mbox{\boldmath$r$}},\,{\mbox{\boldmath$k$}};\,t)}{\partial t}+\Big<\frac{d{\mbox{\boldmath$R$}}_0(t)}{dt}\Big>_{\rm av}\cdot\mbox{\boldmath$\nabla$}_{{\bf r}}f_c({\mbox{\boldmath$r$}},\,{\mbox{\boldmath$k$}};\,t)
\]
\begin{equation}
+\Big<\frac{d{\mbox{\boldmath$K$}_0}(t)}{dt}\Big>_{\rm av}\cdot\mbox{\boldmath$\nabla$}_{{\bf k}}f_c({\mbox{\boldmath$r$}},\,{\mbox{\boldmath$k$}};\,t)
=\left.\frac{\partial f_c({\mbox{\boldmath$r$}},\,{\mbox{\boldmath$k$}};\,t)}{\partial t}\right|_{coll}\ ,
\label{dan-42}
\end{equation}
where ${\mbox{\boldmath$r$}}$ is a three-dimensional position vector, ${\mbox{\boldmath$k$}}$ is a three-dimensional wave vector,
and the term on the right-hand side of this equation corresponds to the collision contribution of electrons with other electrons, impurities, and phonons. Moreover, for conduction-band electrons,
we can define, in a semiclassical way\,\cite{niu}, their group velocity through ${\mbox{\boldmath$v$}}_c({\mbox{\boldmath$k$}})=\mbox{\boldmath$\nabla$}_{\bf k}\varepsilon_c({\mbox{\boldmath$k$}})/\hbar\equiv\langle d{\mbox{\boldmath$R$}_0}(t)/dt\rangle_{\rm av}$,
\color{black}{where $\mbox{\boldmath$R$}_0(t)$ is the center-of-mass position vector.}
\color{black}
Furthermore, we introduce the semiclassical Newton-like force equation\,\cite{niu} for the wave vector of miniband electrons, yielding

\begin{equation}
\hbar\,\Big<\frac{d{\mbox{\boldmath$K$}_0}(t)}{dt}\Big>_{\rm av}={\mbox{\boldmath$F$}}_c({\mbox{\boldmath$k$}},\,t)=-e\left[{\mbox{\boldmath$F$}}(t)+{\mbox{\boldmath$v$}}_c({\mbox{\boldmath$k$}})\times {\mbox{\boldmath$B$}}(t)\right]\ ,
\label{dan-43}
\end{equation}
where \color{black}{$\mbox{\boldmath$K$}_0(t)$ is the center-of-mass wave vector,}
\color{black}
${\mbox{\boldmath$E$}}(t)$ and ${\mbox{\boldmath$B$}}(t)$ are the external electric and magnetic fields, respectively, and ${\mbox{\boldmath$F$}}_c({\mbox{\boldmath$k$}},\,t)$ is the electromagnetic force acting on an electron in the ${\mbox{\boldmath$k$}}$ state.
\medskip

Based on Eq.\,(\ref{dan-42}), the zeroth-order Boltzmann moment equation can be obtained by summing over all the ${\mbox{\boldmath$k$}}$ states on both sides of this equation.
This gives rise to the electron number conservation equation, i.e.,

\begin{equation}
\frac{\partial\rho_c}{\partial t}+\mbox{\boldmath$\nabla$}_{{\bf r}}\cdot{\mbox{\boldmath$J$}}_c=0\ ,
\label{dan-44}
\end{equation}
where the electron number volume density $\rho_c({\mbox{\boldmath$r$}},\,t)$ and particle-number current density ${\mbox{\boldmath$J$}}_c({\mbox{\boldmath$r$}},\,t)$ (per area) are defined by

\begin{equation}
\rho_c({\bf r},\,t)\equiv\frac{2}{{\cal V}}\sum_{{\bf k}}\,f_c({\mbox{\boldmath$r$}},\,{\mbox{\boldmath$k$}};\,t)\ ,
\label{dan-45}
\end{equation}

\begin{equation}
{\mbox{\boldmath$J$}}_c({\mbox{\boldmath$r$}},\,t)\equiv\frac{2}{{\cal V}}\sum_{{\bf k}}\,{\mbox{\boldmath$v$}}_c({\mbox{\boldmath$k$}})\,f_c({\mbox{\boldmath$r$}},\,{\mbox{\boldmath$k$}};\,t)\ ,
\label{dan-46}
\end{equation}
and ${\cal V}$ is the sample volume.
\medskip

For the first-order Boltzmann moment equation, we have to employ the so-called Fermi kinetics. Therefore, we first introduce the {\em relaxation-time approximation} for the electron collision, given by

\begin{equation}
\left.\frac{\partial f_c({\mbox{\boldmath$r$}},\,{\mbox{\boldmath$k$}};\,t)}{\partial t}\right|_{coll}=-\,\frac{f_c({\mbox{\boldmath$r$}},\,{\mbox{\boldmath$k$}};\,t)-f_0[\varepsilon_c({\mbox{\boldmath$k$}}),T;\,u_c]}{\tau_c(k)}\ ,
\label{dan-47}
\end{equation}
where $f_0[\varepsilon_c({\mbox{\boldmath$k$}}),T;\,u_c]=\{\exp[\varepsilon_c({\mbox{\boldmath$k$}})-u_c]/k_BT)]+1\}^{-1}$ is the Fermi function for electrons in thermal-equilibrium states,
$T$ is the lattice temperature, $u_c$ is the chemical potential of electrons in the system, and $\tau_c(k)$ is the energy-relaxation time for electrons in the ${\mbox{\boldmath$k$}}$ state.
The detailed calculation of $\tau_c(k)$ has been presented in Appendix\ \ref{app-1}.
The chemical potential $u_c$ of the system is determined self-consistently by

\begin{equation}
2\,\sum_{{\bf k}}\,f_0[\varepsilon_c({\mbox{\boldmath$k$}}),T;\,u_c]=\int d^3{\mbox{\boldmath$r$}}\,\rho_c({\mbox{\boldmath$r$}},\,t)\equiv\frac{2}{{\cal V}}\sum_{{\bf k}}\,\int d^3{\mbox{\boldmath$r$}}\,f_c({\mbox{\boldmath$r$}},\,{\mbox{\boldmath$k$}};\,t)=N_e\ ,
\label{dan-48}
\end{equation}
where $N_e$ represents the total number of electrons in the system.
Finally, by applying this relaxation-time approximation to the standard Boltzmann transport equation in Eq.\,(\ref{dan-42}), we obtain

\[
f_c({\mbox{\boldmath$r$}},\,{\mbox{\boldmath$k$}};\,t)+\tau_e(T,\,u_c)\,\frac{\partial f_c({\mbox{\boldmath$r$}},\,{\mbox{\boldmath$k$}};\,t)}{\partial t}\approx f_0[\varepsilon_c({\mbox{\boldmath$k$}}),T;\,u_c]
\]
\[
-\frac{\tau_e(T,\,u_c)}{\hbar}\,{\mbox{\boldmath$F$}}_c({\mbox{\boldmath$k$}},\,t)\cdot\mbox{\boldmath$\nabla$}_{{\bf k}}f_0[\varepsilon_c({\mbox{\boldmath$k$}}),T;\,u_c]
-\tau_e(T,\,u_c)\,\mbox{\boldmath$\nabla$}_{{\bf r}}\cdot\left\{{\mbox{\boldmath$v$}}_c({\mbox{\boldmath$k$}})\,f_0[\varepsilon_c({\mbox{\boldmath$k$}}),T;\,u_c]\right\}
\]
\begin{equation}
=f_0[\varepsilon_c({\mbox{\boldmath$k$}}),T;\,u_c]
-\frac{\tau_e(T,\,u_c)}{\hbar}\,{\mbox{\boldmath$F$}}_c({\mbox{\boldmath$k$}},\,t)\cdot\mbox{\boldmath$\nabla$}_{{\bf k}}f_0[\varepsilon_c({\mbox{\boldmath$k$}}),T;\,u_c]\ ,
\label{dan-49}
\end{equation}
where we have used the fact that $T$ is spatially uniform throughout the system and equals the lattice temperature, and the statistically-averaged energy-relaxation time $\tau_e(T,\,u_c)$ is defined by

\begin{equation}
\frac{1}{\tau_e(T,\,u_c)}=\frac{2}{N_e}\sum_{{\bf k}}\,\frac{f_0[\varepsilon_c({\mbox{\boldmath$k$}}),T;\,u_c]}{\tau_c(k)}\ .
\label{dan-50}
\end{equation}
By introducing another inverse momentum-relaxation time tensor $\tensor{\mbox{\boldmath$\tau$}}_p^{-1}$
and using Eq.\,(\ref{dan-43}), we can further write the {\em force-balance equation} for a macroscopic drift velocity $\mbox{\boldmath$v$}_d(t)$, which yields

\[
\frac{d\mbox{\boldmath$v$}_d(t)}{dt}=-\tensor{\mbox{\boldmath$\tau$}}_p^{-1}\cdot\mbox{\boldmath$v$}_d(t)+\tensor{\mbox{\boldmath$\cal M$}}^{-1}\cdot{\mbox{\boldmath$F$}}_e(t)
\]
\begin{equation}
=-\tensor{\mbox{\boldmath$\tau$}_p}^{-1}\cdot\mbox{\boldmath$v$}_d(t)
-e\tensor{\mbox{\boldmath${\cal M}$}}^{-1}\cdot\left[{\mbox{\boldmath$E$}}(t)+\mbox{\boldmath$v$}_d(t)\times{\mbox{\boldmath$B$}}(t)\right]=0\ ,
\label{dan-51}
\end{equation}
where ${\mbox{\boldmath$F$}}_e(t)=-e\left[{\mbox{\boldmath$E$}}(t)+\mbox{\boldmath$v$}_d(t)\times{\mbox{\boldmath$B$}}(t)\right]$ is the macroscopic electromagnetic force, and
the statistically-averaged inverse effective-mass tensor $\tensor{\mbox{\boldmath${{\cal M}}$}}^{-1}$ for conduction-band electrons is given by

\begin{equation}
{\cal M}_{ij}^{-1}(T,\,u_c)=\frac{2}{N_e}\,\sum_{{\bf k}}\,\left[\frac{1}{\hbar^2}\,\frac{\partial^2\varepsilon_c({\mbox{\boldmath$k$}})}{\partial k_i\partial k_j}\right]f_0[\varepsilon_c({\mbox{\boldmath$k$}}),T;\,u_c]\ ,
\label{dan-52}
\end{equation}
and $i,\,j=x,\,y,\,z$. The detailed calculations for the inverse momentum-relaxation time tensor $\tensor{\mbox{\boldmath$\tau$}}_p^{-1}$ in our system can be found in Appendix\ \ref{app-2}.
\color{black}{Moreover, the internal Coulomb force between a pair of electrons will not contribute to this force-balance equation.}
\color{black}
The solution of Eq.\,(\ref{dan-51}) can be formally expressed as

\begin{equation}
\mbox{\boldmath$v$}_d(t)=\tensor{\mbox{\boldmath${\mu}$}}[{\mbox{\boldmath$B$}}(t)]\cdot{\mbox{\boldmath$E$}}(t)\ ,
\label{dan-53}
\end{equation}
where $\tensor{\mbox{\boldmath${\mu}$}}[{\mbox{\boldmath$B$}}(t)]$ is the so-called mobility tensor for conduction-band electrons, 
which also depends on $\tensor{\mbox{\boldmath$\tau$}}_p^{-1}$ and $\tensor{\mbox{\boldmath${\cal M}$}}^{-1}$ in addition to ${\mbox{\boldmath$B$}}(t)$.
The details for calculating the mobility tensor $\tensor{\mbox{\boldmath${\mu}$}}[{\mbox{\boldmath$B$}}(t)]$ are presented in Appendix \ref{app-3}.
Using Eqs.\,(\ref{dan-51}) and (\ref{dan-53}), we can rewrite ${\mbox{\boldmath$F$}}_e(t)=\left(\tensor{\mbox{\boldmath${\cal M}$}}\otimes\tensor{\mbox{\boldmath$\tau$}_p}^{-1}\right)\cdot\left[\tensor{\mbox{\boldmath${\mu}$}}[{\mbox{\boldmath$B$}}(t)]\cdot{\mbox{\boldmath$E$}}(t)\right]$, where
$\tensor{\mbox{\boldmath${\cal M}$}}$ represents the inverse of $\tensor{\mbox{\boldmath${\cal M}$}}^{-1}$.
\medskip

In a similar way, multiplying both sides of Eq.\,(\ref{dan-49}) by ${\mbox{\boldmath$v$}}_c({\mbox{\boldmath$k$}})$ and summing over all the ${\mbox{\boldmath$k$}}$ states afterwards, we get

\[
{\mbox{\boldmath$J$}}_c(t)+\tau_e(T,\,u_c)\,\frac{\partial {\mbox{\boldmath$J$}}_c(t)}{\partial t}=-\tau_e(T,\,u_c)\,\frac{2}{{\cal V}}\,\sum_{{\bf k}}\,{\mbox{\boldmath$v$}}_c({\mbox{\boldmath$k$}})
\left[{\mbox{\boldmath$F$}}_e(t)\cdot{\mbox{\boldmath$v$}}_c({\mbox{\boldmath$k$}})\right]\frac{\partial f_0[\varepsilon_c({\mbox{\boldmath$k$}}),T;\,u_c]}{\partial\varepsilon_c}
\]
\begin{equation}
=-e\tau_e(T,\,u_c)\,\frac{2}{{\cal V}}\,\sum_{{\bf k}}\,{\mbox{\boldmath$v$}}_c({\mbox{\boldmath$k$}})
\left\{\left(\tensor{\mbox{\boldmath${\cal M}$}}\otimes\tensor{\mbox{\boldmath${\tau}$}_p}^{-1}\right)\cdot\left[\tensor{\mbox{\boldmath${\mu}$}}[{\mbox{\boldmath$B$}}(t)]\cdot{\mbox{\boldmath$E$}}(t)\right]\right\}\cdot{\mbox{\boldmath$v$}}_c({\mbox{\boldmath$k$}})\,\left[-\frac{\partial f_0[\varepsilon_c({\mbox{\boldmath$k$}}),T;\,u_c]}{\partial\varepsilon_c}\right]\ .
\label{dan-54}
\end{equation}
From Eq.\,(\ref{dan-54}) we know the particle-number current density $\mbox{\boldmath$J$}_c$ is independent of $\mbox{\boldmath$r$}$. Consequently, from Eq.\,(\ref{dan-44}) we find that the number volume density $\rho_c$
becomes a constant $\rho_0$, determined by

\begin{equation}
\rho_0=\frac{2}{{\cal V}}\sum_{{\bf k}}\,f_0[\varepsilon_c({\mbox{\boldmath$k$}}),T;\,u_c]\ ,
\label{dan-55}
\end{equation}
which determines the chemical potential $u_c$ of the system for fixed $T$. If the external fields are static ones, i.e., ${\mbox{\boldmath$E$}}_0$ and ${\mbox{\boldmath$B$}}_0$, we get the charge current density ${\mbox{\boldmath$J$}}_0$ from Eq.\,(\ref{dan-54})

\begin{equation}
\mbox{\boldmath$J$}_0=-e^2\tau_e\left(\frac{2}{{\cal V}}\right)\sum_{{\bf k}}\,\mbox{\boldmath$v$}_c({\mbox{\boldmath$k$}})
\left\{\left(\tensor{\mbox{\boldmath${\cal M}$}}\otimes\tensor{\mbox{\boldmath${\tau}$}}_p^{-1}\right)\cdot\left[\tensor{\mbox{\boldmath${\mu}$}}[{\mbox{\boldmath$B$}}_0]\cdot{\mbox{\boldmath$E$}}_0)\right]\right\}
\cdot\mbox{\boldmath$v$}_c({\mbox{\boldmath$k$}})\,
\left[-\frac{\partial f_0[\varepsilon_c({\mbox{\boldmath$k$}}),T;\,u_c]}{\partial\varepsilon_c}\right]\ .
\label{dan-56}
\end{equation}
In this case, the elements of the conductivity tensor $\tensor{\mbox{\boldmath${\sigma}$}}({\mbox{\boldmath$B$}}_0)$ can be obtained through
$\sigma_{ij}({\mbox{\boldmath$B$}}_0)={\mbox{\boldmath$J$}}_0\cdot\hat{\mbox{\boldmath$e$}}_i/({\mbox{\boldmath$E$}}_0\cdot\hat{\mbox{\boldmath$e$}}_j)$, where $i,\,j=x,\,y,\,z$ and $\hat{\mbox{\boldmath$e$}}_x,\,\hat{\mbox{\boldmath$e$}}_y,\,\hat{\mbox{\boldmath$e$}}_z$ are three unit vectors
in a position space.
From Eq.\,(\ref{dan-56}), we know that the conductivity tensor depends not only on the mobility tensor, but also on how electrons are distributed within an anisotropic conduction band.
\medskip

As a special case, we consider an isotropic parabolic band structure given by $\varepsilon_c({\mbox{\boldmath$k$}})=\hbar^2k^2/2m^\ast$,
we find from Eq.\,(\ref{dan-52}) that ${\cal M}_{ij}^{-1}=(1/m^\ast)\,\delta_{ij}$ and ${\cal M}_{ij}=m^\ast\,\delta_{ij}$,
as well as $(\tensor{\tau_{p}}^{-1})_{ij}=(1/\tau_p)\,\delta_{ij}$. In this case, from Eq.\,(\ref{dan-71}) we obtain the mobility tensor as

\begin{equation}
\tensor{\mbox{\boldmath${\mu}$}}[{\mbox{\boldmath$B$}}(t)]=-\frac{\mu_0}{1+\mu_0^2B^2}\,
\left[\begin{array}{ccc}
1+\mu_0^2B_x^2 & -\mu_0B_z+\mu_0^2B_xB_y & \mu_0B_y+\mu_0^2B_xB_z\\
\mu_0B_z+\mu_0^2B_yB_x & 1+\mu_0^2B_y^2 & -\mu_0B_x+\mu_0^2B_yB_z\\
-\mu_0B_y+\mu_0^2B_zB_x & \mu_0B_x+\mu_0^2B_zB_y & 1+\mu_0^2B_z^2
\end{array}\right]\ ,
\label{dan-57}
\end{equation}
where $\mu_0=e\tau_p/m^\ast$, ${\mbox{\boldmath$B$}}=\{B_x,\,B_y,\,B_z\}$, and $B^2=B_x^2+B_y^2+B_z^2$. If we further assume ${\mbox{\boldmath$B$}}=0$, Eq.\,(\ref{dan-57}) simply leads to
$\mu_{ij}=-\mu_0\,\delta_{ij}$. In this case, from Eq.\,(\ref{dan-56}) we get the well-known result
${\mbox{\boldmath$J$}}_0=(\rho_0e^2\tau_e/m^\ast)\,{\mbox{\boldmath$E$}}_0$, which implies $\sigma_{ij}=(\rho_0e^2\tau_e/m^\ast)\,\delta_{ij}$.
\medskip
For a $p$-doped semiconductor bulk material, similar equations can be derived for $f_v({\mbox{\boldmath$r$}},\,{\mbox{\boldmath$k$}};\,t)$, $\rho_v({\mbox{\boldmath$r$}},\,t)$ and ${\mbox{\boldmath$J$}}_v(t)$ with replacements of $\varepsilon_c({\mbox{\boldmath$k$}})$, ${\mbox{\boldmath$r$}}_c(t)$,
${\mbox{\boldmath$F$}}_e(t)$, $\tau_c({\mbox{\boldmath$k$}})$, $u_c$, $-e$
by $\varepsilon_v({\mbox{\boldmath$k$}})$, ${\mbox{\boldmath$r$}}_v(t)$, ${\mbox{\boldmath$F$}}_h(t)$, $\tau_v({\mbox{\boldmath$k$}})$, $u_v$, $+e$, respectively.

\section{Surface Boltzmann Moment Equation}

For a semiconductor sheet,
we will also start with the standard semiclassical Boltzmann transport equation for electrons within conduction subbands $\varepsilon_n({\mbox{\boldmath$k$}}_\|)$ of a sheet,
where $n=1,\,2$ for two spin-resolved conduction subbands within the bulk semiconductor bandgap. For this case, the electron distribution function $f_n({\mbox{\boldmath$r$}}_\|,\,{\mbox{\boldmath$k$}}_\|;\,t)$ satisfies

\[
\frac{\partial f_n({\mbox{\boldmath$r$}}_\|,\,{\mbox{\boldmath$k$}}_\|;\,t)}{\partial t}+\Big<\frac{d{\mbox{\boldmath$R$}}_\|(t)}{dt}\Big>_{\rm av}\cdot\mbox{\boldmath$\nabla$}_{{\bf r}_\|}f_n({\mbox{\boldmath$r$}}_\|,\,{\mbox{\boldmath$k$}}_\|;\,t)
\]
\begin{equation}
+\Big<\frac{d{\mbox{\boldmath$K$}}_\|(t)}{dt}\Big>_{\rm av}\cdot\mbox{\boldmath$\nabla$}_{{\bf k}_\|}f_n({\mbox{\boldmath$r$}}_\|,\,{\mbox{\boldmath$k$}}_\|;\,t)
=\left.\frac{\partial f_n({\mbox{\boldmath$r$}}_\|,\,{\mbox{\boldmath$k$}}_\|;\,t)}{\partial t}\right|_{coll}\ ,
\label{dan-142}
\end{equation}
where ${\mbox{\boldmath$r$}}_\|$ is a two-dimensional position vector on the bulk surface, ${\mbox{\boldmath$k$}}_\|$ is a two-dimensional wave vector within the surface plane,
and the term at the right-hand side of this equation corresponds to the collision contribution of electrons with other electrons, impurities, and phonons. Moreover, for conduction-subband electrons,
we can define, in a semiclassical way, their group velocity through ${\mbox{\boldmath$v$}}_n({\mbox{\boldmath$k$}}_\|)=\mbox{\boldmath$\nabla$}_{{\bf k}_\|}\varepsilon_n({\mbox{\boldmath$k$}}_\|)/\hbar\equiv\langle d{\mbox{\boldmath$R$}}_\|(t)/dt\rangle_{\rm av}$.
Furthermore, we introduce the semiclassical Newton-like force equation for the wave vector of miniband electrons, yielding

\begin{equation}
\hbar\,\Big<\frac{d{\mbox{\boldmath$K$}}_\|(t)}{dt}\Big>_{\rm av}\equiv {\mbox{\boldmath$F$}}_n({\mbox{\boldmath$k$}}_\|,\,t)=-e\left[{\mbox{\boldmath$E$}}(t)+{\mbox{\boldmath$v$}}_n({\mbox{\boldmath$k$}}_\|)\times {\mbox{\boldmath$B$}}(t)\right]\ ,
\label{dan-143}
\end{equation}
where ${\mbox{\boldmath$E$}}(t)$ and ${\mbox{\boldmath$B$}}(t)$ are the external electric and magnetic fields, respectively, and ${\mbox{\boldmath$F$}}_n({\mbox{\boldmath$k$}}_\|,\,t)$ is the electromagnetic force acted on an electron in the ${\mbox{\boldmath$k$}}_\|$ state of the
$n$th subband.
\medskip

Based on Eq.\,(\ref{dan-142}), the zeroth-order Boltzmann moment equation can be obtained by summing over all the ${\mbox{\boldmath$k$}}_\|$ states and all the subbands on both sides of this equation.
This gives rise to the electron number conservation equation, i.e.,

\begin{equation}
\frac{\partial n_s}{\partial t}+\mbox{\boldmath$\nabla$}_{{\bf r}_\|}\cdot{\mbox{\boldmath$j$}}_s=0\ ,
\label{dan-144}
\end{equation}
where the surface density of electron number $n_s({\mbox{\boldmath$r$}}_\|,\,t)$ and surface particle-number current density ${\mbox{\boldmath$j$}}_s({\mbox{\boldmath$r$}}_\|,\,t)$ (per length) are defined by

\begin{equation}
n_s({\mbox{\boldmath$r$}}_\|,\,t)\equiv\frac{1}{{\cal A}}\sum_{n,{\bf k}_\|}\,f_n({\mbox{\boldmath$r$}}_\|,\,{\mbox{\boldmath$k$}}_\|;\,t)\ ,
\label{dan-145}
\end{equation}

\begin{equation}
{\mbox{\boldmath$j$}}_s({\mbox{\boldmath$r$}}_\|,\,t)\equiv\frac{1}{{\cal A}}\sum_{n,{\bf k}_\|}\,{\mbox{\boldmath$v$}}_n({\mbox{\boldmath$k$}}_\|)\,f_n({\mbox{\boldmath$r$}}_\|,\,{\mbox{\boldmath$k$}}_\|;\,t)\ ,
\label{dan-146}
\end{equation}
${\cal A}$ is the surface area.
\medskip

For the first-order Boltzmann moment equation, we again have to employ the so-called Fermi kinetics. Therefore, we first introduce the relaxation-time approximation for the electron collision, given by

\begin{equation}
\left.\frac{\partial f_n({\mbox{\boldmath$r$}}_\|,\,{\mbox{\boldmath$k$}}_\|;\,t)}{\partial t}\right|_{coll}=-\,\frac{f_n({\mbox{\boldmath$r$}}_\|,\,{\mbox{\boldmath$k$}}_\|;\,t)-f_0[\varepsilon_n({\mbox{\boldmath$k$}}_\|),T;\,u_c]}{\tau_n(k_\|)}\ ,
\label{dan-147}
\end{equation}
where $f_0[\varepsilon_n({\mbox{\boldmath$k$}}_\|),T;\,u_s]=\{\exp[\varepsilon_n({\mbox{\boldmath$k$}}_\|)-u_s]/k_BT)]+1\}^{-1}$ is the Fermi function for electrons in thermal-equilibrium states,
$T$ is the lattice temperature, $u_s$ is the chemical potential of surface electrons and $\tau_n(k_\|)$ is the energy-relaxation time for electrons in the ${\mbox{\boldmath$k$}}_\|$ state of the $n$th subband.
The surface chemical potential $u_s$ is determined self-consistently by

\begin{equation}
\sum_{n,{\bf k}_\|}\,f_0[\varepsilon_n({\mbox{\boldmath$k$}}_\|),T;\,u_s]=\int d^2{\mbox{\boldmath$r$}}_\|\,n_s({\mbox{\boldmath$r$}}_\|,\,t)\equiv\frac{1}{{\cal A}}\sum_{n,{\bf k}_\|}\,\int d^2{\mbox{\boldmath$r$}}_\|\,f_n({\mbox{\boldmath$r$}}_\|,\,{\mbox{\boldmath$k$}}_\|;\,t)=N_s\ ,
\label{dan-148}
\end{equation}
where $N_s=n_0{\cal A}$ represents the total number of surface electrons for each spin \color{black}{and $n_0$ is the areal density for surface electrons}.
\color{black}
Finally, by applying this relaxation-time approximation to the standard Boltzmann transport equation in Eq.\,(\ref{dan-142}), we obtain

\[
f_n({\mbox{\boldmath$r$}}_\|,\,{\mbox{\boldmath$k$}}_\|;\,t)+\tau_s(T,\,u_s)\,\frac{\partial f_n({\mbox{\boldmath$r$}}_\|,\,{\mbox{\boldmath$k$}}_\|;\,t)}{\partial t}\approx f_0[\varepsilon_n({\mbox{\boldmath$k$}}_\|),T;\,u_s]
\]
\[
-\frac{\tau_s(T,\,u_s)}{\hbar}\,{\mbox{\boldmath$F$}}_n({\mbox{\boldmath$k$}}_\|,\,t)\cdot\mbox{\boldmath$\nabla$}_{{\bf k}_\|}f_0[\varepsilon_n({\mbox{\boldmath$k$}}_\|),T;\,u_s]
-\tau_s(T,\,u_s)\,\mbox{\boldmath$\nabla$}_{{\bf r}_\|}\cdot\left\{{\mbox{\boldmath$v$}}_n({\mbox{\boldmath$k$}}_\|)\,f_0[\varepsilon_n({\mbox{\boldmath$k$}}_\|),T;\,u_s]\right\}
\]
\begin{equation}
=f_0[\varepsilon_n({\mbox{\boldmath$k$}}_\|),T;\,u_s]
-\frac{\tau_s(T,\,u_s)}{\hbar}\,{\mbox{\boldmath$F$}}_n({\mbox{\boldmath$k$}}_\|,\,t)\cdot\mbox{\boldmath$\nabla$}_{{\bf k}_\|}f_0[\varepsilon_n({\mbox{\boldmath$k$}}_\|),T;\,u_s]\ ,
\label{dan-149}
\end{equation}
where we have used the fact that $T$ is uniform throughout the system and equals the lattice temperature, and the statistically-averaged surface energy-relaxation time $\tau_s(T,\,u_s)$ is defined by

\begin{equation}
\frac{1}{\tau_s(T,\,u_s)}=\frac{1}{N_s}\sum_{n,{\bf k}_\|}\,\frac{f_0[\varepsilon_n({\mbox{\boldmath$k$}}_\|),T;\,u_s]}{\tau_n(k_\|)}\ .
\label{dan-150}
\end{equation}
By introducing another inverse surface momentum-relaxation time tensor $\tensor{\mbox{\boldmath$\tau$}}_{sp}^{-1}$
and using Eq.\,(\ref{dan-143}), we can further write the force-balance equation for the macroscopic surface drift velocity $\mbox{\boldmath$v$}_s(t)$, which yields

\[
\frac{d\mbox{\boldmath$v$}_s(t)}{dt}=-\tensor{\mbox{\boldmath$\tau$}}_{sp}^{-1}\cdot\mbox{\boldmath$v$}_s(t)+\tensor{\mbox{\boldmath$\cal M$}}_s^{-1}\cdot{\mbox{\boldmath$F$}}_s(t)
\]
\begin{equation}
=-\tensor{\mbox{\boldmath$\tau$}}_{sp}^{-1}\cdot\mbox{\boldmath$v$}_s(t)
-e\tensor{\mbox{\boldmath${\cal M}$}}_s^{-1}\cdot\left[{\mbox{\boldmath$E$}}(t)+\mbox{\boldmath$v$}_s(t)\times{\mbox{\boldmath$B$}}(t)\right]=0\ ,
\label{dan-151}
\end{equation}
where the macroscopic surface electromagnetic force is ${\mbox{\boldmath$F$}}_s(t)=-e\left[{\mbox{\boldmath$E$}}(t)+\mbox{\boldmath$v$}_s(t)\times{\mbox{\boldmath$B$}}(t)\right]$, and
the statistically-averaged inverse effective-mass tensor for surface electrons is given by

\begin{equation}
({\cal M}^{-1}_s)_{ij}=\frac{1}{N_s}\,\sum_{n,{\bf k}_\|}\,\left[\frac{1}{\hbar^2}\,\frac{\partial^2\varepsilon_n({\mbox{\boldmath$k$}}_\|)}{\partial k_{i\|}\partial k_{j\|}}\right]f_0[\varepsilon_n({\mbox{\boldmath$k$}}_\|),T;\,u_s]\ ,
\label{dan-152}
\end{equation}
and $i,\,j=x,\,y$.
The solution of Eq.\,(\ref{dan-151}) can be formally written as

\begin{equation}
\mbox{\boldmath$v$}_s(t)=\tensor{\mbox{\boldmath${\mu}$}}_s[{\mbox{\boldmath$B$}}(t)]\cdot{\mbox{\boldmath$E$}}(t)\ ,
\label{dan-153}
\end{equation}
where $\tensor{\mbox{\boldmath${\mu}$}}_s[{\mbox{\boldmath$B$}}(t)]$ is the so-called mobility tensor for surface electrons, which also depends on $\tensor{\mbox{\boldmath$\tau$}}_{sp}^{-1}$ and $\tensor{\mbox{\boldmath${\cal M}$}}_s^{-1}$ in addition to ${\mbox{\boldmath$B$}}(t)$.
Using Eqs.\,(\ref{dan-151}) and (\ref{dan-153}), we can rewrite ${\mbox{\boldmath$F$}}_s(t)=\left(\tensor{\mbox{\boldmath${\cal M}$}}_s\otimes\tensor{\mbox{\boldmath$\tau$}_p}^{-1}\right)\cdot\left[\tensor{\mbox{\boldmath${\mu}$}}_s[{\mbox{\boldmath$B$}}(t)]\cdot{\mbox{\boldmath$E$}}(t)\right]$, where
$\tensor{\mbox{\boldmath${\cal M}$}}_s$ represents the inverse of $\tensor{\mbox{\boldmath${\cal M}$}}_s^{-1}$.
\medskip

In a similar way, multiplying both sides of Eq.\,(\ref{dan-149}) by ${\mbox{\boldmath$v$}}_n({\mbox{\boldmath$k$}}_\|)$ and summing over all the ${\mbox{\boldmath$k$}}_\|$ states and all the subbands afterwards, we get

\[
{\mbox{\boldmath$j$}}_s(t)+\tau_s(T,\,u_s)\,\frac{\partial {\mbox{\boldmath$j$}}_s(t)}{\partial t}=-\tau_s(T,\,u_s)\,\frac{1}{{\cal A}}\,\sum_{n,{\bf k}_\|}\,{\mbox{\boldmath$v$}}_n({\mbox{\boldmath$k$}}_\|)
\left[{\mbox{\boldmath$F$}}_s(t)\cdot{\mbox{\boldmath$v$}}_n({\mbox{\boldmath$k$}}_\|)\right]\frac{\partial f_0[\varepsilon_n({\mbox{\boldmath$k$}}_\|),T;\,u_s]}{\partial\varepsilon_n}
\]
\begin{equation}
=e\tau_s(T,\,u_s)\,\frac{1}{{\cal A}}\,\sum_{n,{\bf k}_\|}\,{\mbox{\boldmath$v$}}_n({\mbox{\boldmath$k$}_\|})
\left\{\left(\tensor{\mbox{\boldmath${\cal M}$}}_s\otimes\tensor{\mbox{\boldmath${\tau}$}}_{sp}^{-1}\right)\cdot\left[\tensor{\mbox{\boldmath${\mu}$}}_s[{\mbox{\boldmath$B$}}(t)]\cdot{\mbox{\boldmath$E$}}(t)\right]\right\}\cdot{\mbox{\boldmath$v$}}_n({\mbox{\boldmath$k$}}_\|)\,\left[-\frac{\partial f_0[\varepsilon_n({\mbox{\boldmath$k$}}_\|),T;\,u_s]}{\partial\varepsilon_n}\right]\ .
\label{dan-154}
\end{equation}
From Eq.\,(\ref{dan-154}) we know the surface particle-number current density ${\mbox{\boldmath$j$}}_s$ is independent of ${\mbox{\boldmath$r$}}_\|$. As a result, from Eq.\,(\ref{dan-144}) we find the surface number areal density $n_s$
becomes a constant $n_0$, determined by

\begin{equation}
n_0=\frac{1}{{\cal A}}\sum_{n,{\bf k}_\|}\,f_0[\varepsilon_n({\mbox{\boldmath$k$}}_\|),T;\,u_s]\ ,
\label{dan-155}
\end{equation}
which determines the surface chemical potential $u_s$ for fixed $T$. If the external fields are static ones, i.e., ${\mbox{\boldmath$E$}}_0$ and ${\mbox{\boldmath$B$}}_0$, we get the surface charge-current density ${\mbox{\boldmath$j$}}_0$ from Eq.\,(\ref{dan-154})

\begin{equation}
{\mbox{\boldmath$j$}}_0=e^2\tau_s\left(\frac{1}{{\cal A}}\right)\sum_{n,{\bf k}_\|}\,{\mbox{\boldmath$v$}}_n({\mbox{\boldmath$k$}}_\|)
\left\{\left(\tensor{\mbox{\boldmath${\cal M}$}}_s\otimes\tensor{\mbox{\boldmath${\tau}$}}_{sp}^{-1}\right)\cdot\left[\tensor{\mbox{\boldmath${\mu}$}}_s[{\mbox{\boldmath$B$}}_0]\cdot{\mbox{\boldmath$E$}}_0)\right]\right\}\cdot{\mbox{\boldmath$v$}}_n({\mbox{\boldmath$k$}}_\|)\,
\left[-\frac{\partial f_0[\varepsilon_n({\mbox{\boldmath$k$}}_\|),T;\,u_s]}{\partial\varepsilon_c}\right]\ .
\label{dan-156}
\end{equation}
In this case, the elements of the conductivity tensor $\tensor{\mbox{\boldmath${\sigma}$}}({\mbox{\boldmath$B$}}_0)$ can be obtained through
$\sigma_{ij}({\mbox{\boldmath$B$}}_0)={\mbox{\boldmath$j$}}_0\cdot\hat{\mbox{\boldmath$e$}}_i/({\mbox{\boldmath$E$}}_0\cdot\hat{\mbox{\boldmath$e$}}_j)$, where $i,\,j=x,\,y$ and $\hat{\mbox{\boldmath$e$}}_x,\,\hat{\mbox{\boldmath$e$}}_y$ are the unit vectors.
From Eq.\,(\ref{dan-156}), we know that the conductivity tensor not only depends on the mobility tensor, but also depends on how electrons are distributed within anisotropic conduction subbands.

\section{Coulomb Effect on Surface Conductivity}

From Eqs.\,(\ref{new-4}) and (\ref{new-6}), we find the total inverse momentum-relaxation-time tensor $\tensor{\mbox{\boldmath${\tau}$}}_{sp}^{-1}=\tensor{\mbox{\boldmath${\tau}$}}_{s,i}^{-1}+\tensor{\mbox{\boldmath${\tau}$}}_{s,ph}^{-1}$
in Eq.\,(\ref{dan-156}) through

\begin{equation}
\tensor{\mbox{\boldmath${\tau}$}}_{s,i}^{-1}=-\frac{2\sigma_i}{n_0{\cal A}}\,\sum_{{\bf q}_\|}\,\left|U_i(q_\|)\right|^2\left\{\frac{\partial{\rm Im}\left[\Pi_s(q_\|,\omega)\right]}
{\partial\omega}\right\}_{\omega=\Gamma_0}
\left\{\tensor{\mbox{\boldmath${\cal M}$}}_s^{-1}\otimes\left[{\mbox{\boldmath$q$}}_\|\otimes{\mbox{\boldmath$q$}}_\|^T\right]\right\}\ ,
\label{new-7}
\end{equation}
and

\[
\tensor{\mbox{\boldmath${\tau}$}}_{s,ph}^{-1}=-\frac{4}{n_0{\cal A}}\,\sum_{{\bf q}_\|,\lambda}\,\left|C_{q_\|\lambda}\right|^2\,\left\{\frac{\partial{\rm Im}\left[\Pi_s(q_\|,\omega)\right]}{\partial\omega}\right\}_{\omega=\omega_{q_\|\lambda}}
\]
\[
\times\left[N_0(\omega_{q_\|\lambda},T)-N_0(\omega_{q_\|\lambda}+\mbox{\boldmath$q$}_\|\cdot\mbox{\boldmath$v$}_d,T_e)\right]\,\left\{\tensor{\mbox{\boldmath${\cal M}$}}_s^{-1}\otimes\left[{\mbox{\boldmath$q$}}_\|\otimes{\mbox{\boldmath$q$}}_\|^T\right]\right\}
\]
\[
-\frac{4}{n_0{\cal A}}\,\sum_{{\bf q}_\|}\,\left|C_{q_\|}\right|^2\,\left\{\frac{\partial{\rm Im}\left[\Pi_s(q_\|,\omega)\right]}{\partial\omega}\right\}_{\omega=\omega_{\rm LO}}
\]
\begin{equation}
\times\left[N_0(\omega_{\rm LO},T)-N_0(\omega_{\rm LO}+\mbox{\boldmath$q$}_\|\cdot\mbox{\boldmath$v$}_d,T_e)\right]\,\left\{\tensor{\mbox{\boldmath${\cal M}$}}_s^{-1}\otimes\left[{\mbox{\boldmath$q$}}_\|\otimes{\mbox{\boldmath$q$}}_\|^T\right]\right\}
\label{new-8}
\end{equation}
with

\[
\left[{\mbox{\boldmath$q$}}_\|\otimes{\mbox{\boldmath$q$}}_\|^T\right]\equiv\left[\begin{array}{cc}
q_x^2 & q_xq_y\\
q_yq_x & q_y^2\\
\end{array}\right]\ ,
\]
where $\Gamma_0$ is the inverse of particle lifetime due to vertex correction, \color{black}{$\sigma_i$ is the areal density of impurities, $\omega_{q_\|\lambda}$ and $\omega_{\rm LO}$ are the frequencies of 
acoustic and longitudinal-optical phonons, $N_0(\omega,T)=[\exp(\hbar\omega/k_BT)-1]^{-1}$ is the Bose function for thermal-equilibrium phonons, and $T_e$ is the hot-electron temperature due to inelastic phonon scatterings.}
\color{black}
Here, we assume that only the lowest subband of surface electrons is occupied, and
the imaginary part of the {\em screened} polarization function, ${\rm Im}\left[\Pi_s(q_\|,\omega)\right]$, is given by

\begin{equation}
{\rm Im}\left[\Pi_s(q_\|,\omega)\right]=\frac{{\rm Im}\left[\Pi_s^{(0)}(q_\|,\omega)\right]}{\left\{1-v_s(q_\|)\,{\rm Re}\left[\Pi_s^{(0)}q_\|,\omega)\right]\right\}^2
+\left\{v_s(q_\|)\,{\rm Im}\left[\Pi_s^{(0)}(q_\|,\omega)\right]\right\}^2}\ ,
\label{new-9}
\end{equation}
where the denominator represents the screening effect, $v_s(q_\|)=(e^2/2\epsilon_0\epsilon_bq_\|)\exp(-q_\|\delta_s)$ is the two-dimensional Fourier transform of a bare Coulomb potential, 
\color{black}{$\epsilon_b$ is the dielectric constant of the host material} 
\color{black}
and $\delta_s$ is the thickness of the surface layer.
Moreover, the {\em bare} polarization function $\Pi_s^{(0)}(q_\|,\omega)$ in Eq.\,(\ref{new-9}) is calculated within the random-phase approximation as

\begin{equation}
\Pi_s^{(0)}(q_\|,\omega)=\frac{2}{{\cal A}}\,\sum_{\gamma,\gamma'=\pm 1}\,\sum_{{\bf k}_\|}\,{\cal F}_{\gamma,\gamma'}(\mbox{\boldmath$k$}_\|,\mbox{\boldmath$q$}_\|)\,\frac{f^s_{\gamma,{\bf k}_\|}-f^s_{\gamma',{\bf k}_\|+{\bf q}_\|}}{\hbar\omega+i0^+-\varepsilon^s_{\gamma',{\bf k}_\|+{\bf q}_\|}+\varepsilon^s_{\gamma,{\bf k}_\|}}\ ,
\label{new-10}
\end{equation}
where the overlapping factor for zero-bandgap is given by

\[
{\cal F}_{\gamma,\gamma'}(\mbox{\boldmath$k$}_\|,\mbox{\boldmath$q$}_\|)=\frac{1}{2}\left[1+\gamma\gamma'\,\frac{\mbox{\boldmath$k$}_\|\cdot(\mbox{\boldmath$k$}_\|+\mbox{\boldmath$q$}_\|)}{|\mbox{\boldmath$k$}_\|||\mbox{\boldmath$k$}_\|+\mbox{\boldmath$q$}_\||}\right]\ ,
\]
$\varepsilon^s_{\gamma,{\bf k}_\|}=\gamma\varepsilon^s_{{\bf k}_\|}$
and $f^s_{\gamma,{\bf k}_\|}=f_0[\varepsilon^s_{\gamma,{\bf k}_\|},T_e;u_s]=\left\{1+\exp\left[(\gamma\varepsilon^s_{{\bf k}_\|}-u_s)/k_BT_e\right]\right\}^{-1}$ is the Fermi-Dirac function for thermal-equilibrium surface electrons at
an elevated temperature $T_e$.
\medskip

\color{black}{Let us first consider the case with a zero bandgap, i.e., $\Delta_0=0$.}
\color{black}
For $T_e=0$ and in the long-wavelength limit ($q_\|\to 0$), we obtain\,\cite{r4} from Eq.\,(\ref{new-10})

\begin{equation}
{\rm Re}\left[\Pi_s^{(0)}(q_\|,\omega)\right]=\frac{q_\|^2}{4\pi\hbar\omega}
\left[\frac{4E^s_F}{\hbar\omega}+\ln\left|\frac{2E_F^s-\hbar\omega}{2E_F^s+\hbar\omega}\right|\,\right]\ ,
\label{new-11}
\end{equation}

\begin{equation}
{\rm Im}\left[\Pi_s^{(0)}(q_\|,\omega)\right]=-\frac{q_\|^2}{4\hbar\omega}\,\Theta(\hbar\omega-2E_F^s)\ ,
\label{new-12}
\end{equation}
where $\Theta(x)$ is a unit-step function, $E_F^s=\hbar v_Fk^s_F$ and $k^s_F=\sqrt{4\pi n_0}$.
On the other hand, in the high-temperature $k_BT_e\gg E_F^s,\,\hbar\omega$ and long-wavelength $q_\|\to 0$ limits, we arrive at\,\cite{r4}

\begin{equation}
{\rm Re}\left[\Pi_s^{(0)}(q_\|,\omega)\right]=\frac{2\ln 2}{\pi}\left(\frac{q_\|^2}{\hbar^2\omega^2}\right)k_BT_e\left[1+\frac{(E_F^s)^4}{128(\ln 2)^3(k_BT_e)^4}\right]\ ,
\label{new-13}
\end{equation}

\begin{equation}
{\rm Im}\left[\Pi_s^{(0)}(q_\|,\omega)\right]=-\frac{q_\|^2}{16k_BT_e}\left[1-\frac{\hbar^2\omega^2}{48(k_BT_e)^2}\right]\ .
\label{new-14}
\end{equation}
For $T_e\neq 0$ but $k_BT_e\ll E_F^s,\,\hbar\omega$, the zero-temperature results in Eqs.\,(\ref{new-11}) and (\ref{new-12}) can be formally generalized to\,\cite{r4}

\begin{equation}
{\rm Re}\left[\Pi_s^{(0)}(q_\|,\omega)\right]=\frac{q_\|^2}{4\pi\hbar\omega}
\left[\frac{4u_s(T_e)}{\hbar\omega}+\ln\left|\frac{2u_s(T_e)-\hbar\omega}{2u_s(T_e)+\hbar\omega}\right|\right]\ ,
\label{new-15}
\end{equation}

\begin{equation}
{\rm Im}\left[\Pi_s^{(0)}(q_\|,\omega)\right]=-\frac{q_\|^2}{8\hbar\omega}\left[1+\tanh\left(\frac{\hbar\omega-2u_s(T_e)}{4k_BT_e}\right)\right]
\label{new-16}
\end{equation}
with \color{black}{a chemical potential at finite temperatures}
\color{black}
\[
u_s(T_e)\approx E_F^s\left[1-\frac{\pi^2}{6}\left(\frac{k_BT_e}{E_F^s}\right)^2\right]\ .
\]
\medskip

If we further consider a {\em gaped and undoped} subband for surface electrons with an energy gap $\Delta_0$ and $E_F^s\to 0$, then we acquire the generalized overlapping factor

\[
{\cal F}_{\gamma,\gamma'}(\mbox{\boldmath$k$}_\|,\mbox{\boldmath$q$}_\|)=\frac{1}{2}\left[1+\gamma\gamma'\,\frac{(\hbar v_F)^2\mbox{\boldmath$k$}_\|\cdot(\mbox{\boldmath$k$}_\|+\mbox{\boldmath$q$}_\|)+\Delta^2_0}
{\sqrt{\hbar^2v_F^2|\mbox{\boldmath$k$}_\||^2+\Delta_0^2}\,\sqrt{\hbar^2v_F^2|\mbox{\boldmath$k$}_\|+\mbox{\boldmath$q$}_\||^2+\Delta_0^2}}\right]\ .
\]
Moreover, Eq.\,(\ref{new-10}) under the condition of $k_BT_e\gg\Delta_0,\,\hbar\omega$ turns into\,\cite{r4}

\begin{equation}
{\rm Re}\left[\Pi_s^{(0)}(q_\|,\omega)\right]=\frac{4k_BT_eq_\|^2}{\pi\hbar^2\omega^2}\left\{2\ln 2-\left(\frac{\Delta_0}{k_BT_e}\right)^2\left[C_0-\ln\left(\frac{\Delta_0}{2k_BT_e}\right)\right]\right\}\ ,
\label{new-17}
\end{equation}

\begin{equation}
{\rm Im}\left[\Pi_s^{(0)}(q_\|,\omega)\right]=-\frac{q_\|^2}{16k_BT_e}\left(1-\frac{\Delta_0}{\hbar\omega}\right)\ ,
\label{new-18}
\end{equation}
\medskip
where $C_0\approx 0.79$ is a constant.
\medskip

Firstly, let us consider only the impurity scattering at low temperatures. We know from Eq.\,(\ref{new-7}) that $\tensor{\mbox{\boldmath${\tau}$}}_{sp}^{-1}$ becomes diagonal and its identical diagonal element $1/\tau_{sp}$ is given by

\begin{equation}
\frac{1}{\tau_{sp}}=-\frac{\sigma_i}{2\pi n_0m^\ast}\left(\frac{Z^\ast e^2}{2\epsilon_0\epsilon_b}\right)^2\int\limits_0^{1/\delta_s} dq_\|\,q_\|\left\{\frac{\partial{\rm Im}\left[\Pi_s(q_\|,\omega)\right]}
{\partial\omega}\right\}_{\omega=\Gamma_0}\ .
\label{new-21}
\end{equation}
By making use of the results in Eqs.\,(\ref{new-17}) and (\ref{new-18}), Eq.\,(\ref{new-21}) for impurity scattering leads to
\[
\frac{1}{\tau_{sp}}\approx\frac{v_F^2}{128\pi\hbar\Gamma^2_0\delta_s^2k_BT^*}\left(\frac{\sigma_i}{n_0}\right)
\left(\frac{Z^*e^2}{2\epsilon_0\epsilon_b\delta_s}\right)^2\left(\frac{T^*}{T_e}\right)
\]
\begin{equation}
\times\left\{1-\frac{128}{5\pi\hbar\Gamma_0}\left(\frac{e^2}{2\epsilon_0\epsilon_b\delta_s}\right)\left(1-\frac{3k_BT^*}{\hbar\Gamma_0}\right)
\left(\frac{T^*}{T_e}\right)\left[C_0+\ln\left(\frac{T_e}{T^*}\right)-\frac{\ln2}{2}\left(\frac{T_e}{T^*}\right)^2\right]\right\}\ ,
\label{new-22}
\end{equation}
which satisfies \color{black}{$1/\tau_{e,h}\propto 1/n_0\sim (L_A-L_0)^{-1}$,} 
\color{black}
where we have used $1/m_s^*=v_F^2/\Delta_0$ and defined  $T^*=\Delta_0/2k_B$.
\medskip

The results for phonon scattering can be obtained in a similar way.
Furthermore, we find from Eq.\,(\ref{dan-156}) that

\begin{equation}
{\mbox{\boldmath$j$}}_0=\left(\frac{e^2v_F^2\tau_s}{\Delta_0\tau_{sp}}\right)\frac{1}{\cal A}\sum_{{\bf k}_\|}\,{\mbox{\boldmath$v$}}^s_{{\bf k}_\|}
\left\{\tensor{\mbox{\boldmath${\cal I}$}}_0\cdot\left[\tensor{\mbox{\boldmath${\mu}$}}_s[{\mbox{\boldmath$B$}}_0]\cdot{\mbox{\boldmath$E$}}_0)\right]\right\}\cdot{\mbox{\boldmath$v$}}^s_{{\bf k}_\|}\,
\left[-\frac{\partial f_0[\varepsilon^s_{{\bf k}_\|},T_e;\,u_s]}{\partial\varepsilon^s_{{\bf k}_\|}}\right]\ ,
\label{new-23}
\end{equation}
where \color{black}{${\mbox{\boldmath$v$}}^s_{{\bf k}_\|}=\hbar v_F^2{\mbox{\boldmath$k$}}_\|/\Delta_0=\hbar v_F^2{\mbox{\boldmath$k$}}_\|/2k_BT^*$} 
\color{black}
is the group velocity of surface electrons and
$\tau_s(T_e,u_s)$ is the statistically averaged energy-relaxation time $\tau_s({\mbox{\boldmath${k}$}}_\|)$.
\medskip

Secondly, including the {\em screened} pair scattering of surface electrons, we get $1/\tau^s_{\rm pair}(T,u_s)$ from Eqs.\,(\ref{a-1})-(\ref{a-3}), as well as from Eqs.\,(\ref{new-17}) and (\ref{new-18}), yielding

\[
\frac{1}{\tau^s_{\rm pair}(T_e,u_s)}=\frac{1}{n_0{\cal A}}\sum_{{\bf k}_\|}\,\frac{f^s_{{\bf k}_\|}}{\tau^s_{\rm pair}({\mbox{\boldmath$k$}}_\|)}
\approx\frac{1}{16\pi^4\hbar n_0}\left(\frac{e^2}{2\epsilon_0\epsilon_b}\right)^2
\]
\[
\times\int\limits_{q_0}^{1/\delta_s}
\frac{dq_\|}{q_\|}\,
\left\{1-q_\|\left(\frac{e^2}{2\epsilon_0\epsilon_b}\right)\frac{32k_BT^*}{\pi\hbar^2\Gamma_0^2}\left(\frac{T^*}{T_e}\right)
\left[C_0+\ln\left(\frac{T_e}{T^*}\right)-\frac{\ln2}{2}\left(\frac{T_e}{T^*}\right)^2\right]\right\}
\]
\[
\times\int d^2{\mbox{\boldmath$k$}}_\|\,f^s_{{\bf k}_\|}\int d^2{\mbox{\boldmath$k$}}'_\|
\left[f^s_{{\bf k}'_\|}(1-f^s_{{\bf k}_\|-{\bf q}_\|})(1-f^s_{{\bf k}'_\|+{\bf q}_\|})
+f^s_{{\bf k}_\|-{\bf q}_\|}f^s_{{\bf k}'_\|+{\bf q}_\|}(1-f^s_{{\bf k}'_\|})\right]
\]
\begin{equation}
\times\frac{\Gamma_0/\pi}
{(\varepsilon^s_{{\bf k}_\|}+\varepsilon^s_{{\bf k}'_\|}-\varepsilon^s_{{\bf k}_\|-{\bf q}_\|}-\varepsilon^s_{{\bf k}'_\|+{\bf q}_\|})^2+\Gamma_0^2}\ ,
\label{new-24}
\end{equation}
where 

\begin{equation}
f^s_{{\bf k}_\|}\approx \frac{2\pi\hbar^2v_F^2\,n_0}{(k_BT_e)^2(1+\Delta_0/k_BT_e)}\,\exp\left(-\frac{\varepsilon^s_{{\bf k}_\|}-\Delta_0}{k_BT_e}\right)\ ,
\label{new-26}
\end{equation}
\color{black}{$n_0=(m_s^*/2\pi\hbar^2)E^s_F=(\Delta_0/2\pi\hbar^2v_F^2)E^s_F=(k_BT^*/\pi\hbar^2v_F^2)E^s_F\sim\alpha_0(L_A-L_0)$} 
\color{black}
with $L_A$ as the acceptor-layer thickness,
$\gamma\equiv 1$ is taken and $q_0=\Gamma_0/\hbar v_F$ is a cutoff for $q_\|\to 0$. Here, pair scattering of bulk electrons will 
lead to reduction of total conductivity. Furthermore, $1/\tau^s_{\rm pair}(T_e,u_s)$ has its density dependence of both $\sim n_0$ and $\sim n_0^2$. 
In principle, bulk electrons can also screen impurity scattering of 
surface electrons, but it becomes insignificant due to large separation between the surface layer and the center of the film.
\medskip

Finally, by using Eq.\,(\ref{app-23}) the total conductivity is calculated as

\[
\tensor{\mbox{\boldmath${\sigma}$}}_{tot}({\mbox{\boldmath${B}$}})=e\,\tensor{\mbox{\boldmath${\mu}$}}^\|_{v}({\mbox{\boldmath${B}$}})N_AA_h\left[(L_A-W_p)+\int^{W_p}_{0} dz\,\exp\left(-\frac{\beta e\bar{\mu}_hN_A}{2\epsilon_0\epsilon_rD_h}\,z^2\right)\right]
-e\,\tensor{\mbox{\boldmath${\mu}$}}^\|_{c}({\mbox{\boldmath${B}$}})N_DA_e
\]
\begin{equation}
\times\left[(L_D-W_n)+\int^{W_n}_{0} dz\,\exp\left(-\frac{\beta e\bar{\mu}_eN_D}{2\epsilon_0\epsilon_rD_e}\,z^2\right)\right]+e\,\tensor{\mbox{\boldmath${\mu}$}}^{\pm}_s({\mbox{\boldmath${B}$}})\,\color{black}{\left(\frac{\alpha_0\Delta_0}{2\pi\hbar^2v_F^2}\right)\left(L_A-L_0\right)A_s\ ,}
\label{new-19}
\end{equation}
\color{black}
where $A_s=\tau_s/\tau_{sp}$ and $A_{e,h}=\tau_{e,h}/\tau_{p(e,h)}$. Here, the surface mobility is given by

\begin{equation}
\tensor{\mbox{\boldmath${\mu}$}}_s({\mbox{\boldmath${B}$}})=\frac{\mu_1}{1+\mu^2_1B^2}\,
\left[\begin{array}{cc}
1 & \mu_1B\\
-\mu_1B & 1\\
\end{array}\right]\ ,
\label{new-20}
\end{equation}
and \color{black}{$\mu_1=e\tau_{sp}v^2_F/\Delta_0=e\tau_{sp}v^2_F/2k_BT^*$.} 
\color{black}
For weak magnetic field, we have $\mu_1B\ll 1$, $\mu_{xx}=\mu_{yy}=\mu_1$ and $\mu_{xy}=-\mu_{yx}=\mu_1^2B$.
As ${\mbox{\boldmath${B}$}}\to 0$, we find from Eqs.\,(\ref{new-19}) and (\ref{new-20}) that the change of the total conductivity due to the screened pair scattering of surface electrons is given by

\[
\delta\sigma_{tot}(T_e,u_s)\equiv\sigma_{tot}(T_e,u_s)-\sigma_{tot}^{(0)}(T_e,u_s)
\]
\begin{equation}
=-\mu_0^s\left(\frac{\alpha_0\Delta_0}{2\pi\hbar^2v_F^2}\right)(L_A-L_0)\left[\frac{\tau_0^s(T_e,u_s)}{\tau_0^s(T_e,u_s)+\tau^{s}_{\rm pair}(T_e,u_s)}\right]\approx -\sigma_0^s\left[\frac{\tau_0^s(T_e,u_s)}{\tau^{s}_{\rm pair}(T_e,u_s)}\right]\ ,
\label{new-25}
\end{equation}
where \color{black}{$\mu_0^s=e\tau_0^sv^2_F/\Delta_0=e\tau_0^sv^2_F/2k_BT^*$,}
\color{black}
$\sigma_0^s$ and $\tau_0^s$ 
are the mobility, conductivity and energy-relaxation time, respectively, of surface electrons in the absence of electron-electron interaction (EEI).
\medskip

Therefore, from Eq.\,(\ref{new-25}) we know \color{black}{$\delta\sigma_{tot}(T_e,u_s)\propto\sigma_0^s\sim (L_A-L_0)$.} 
\color{black}
Meanwhile, we also find the ratio \color{black}{$\tau_0^s/\tau^s_{\rm pair}\propto n_0\sim (L_A-L_0)$,}
\color{black}
as can be seen from Eqs.\,(\ref{new-24}) and (\ref{a-2}).
Although the screening due to electron-electron interaction can weaken the impurity scattering and increases the mobility, the conductivity is not affected by the momentum-relaxation time $\tau_{sp}$ of surface electrons.
Even for two-dimensional electron gases in a quantum well, where they acquire a static dielectric function\,\cite{r5} $\epsilon_s(q_\|)\equiv\epsilon(q_\|,\omega=0)=1+q_s/q_\|$ with a Thomas-Fermi screening length $1/q_s$, 
the screened impurity scattering can also increase their conductivity.

\newpage

\section{Energy-Relaxation Time}
\label{app-1}

By using the detailed-balance condition, the energy-relaxation time $\tau_c(k)$ initially introduced in Eq.\,(\ref{dan-47}) can be calculated according to\,\cite{r1}

\begin{equation}
\frac{1}{\tau_c(k)}={\cal W}_{\rm in}(k)+{\cal W}_{\rm out}(k)\ ,
\label{a-1}
\end{equation}
where the scattering-in rate for electrons in the final ${\mbox{\boldmath$k$}}$-state is

\[
{\cal W}_{\rm in}(k)=n_i\,\frac{2\pi}{\hbar{\cal V}}\,\sum_{{\bf q}}\,\left|U_i(q)\right|^2\,
\left[f_{{\bf k}-{\bf q}}\,\delta(\varepsilon_{\bf k}-\varepsilon_{{\bf k}-{\bf q}})+f_{{\bf k}+{\bf q}}\,
\delta(\varepsilon_{\bf k}-\varepsilon_{{\bf k}+{\bf q}})\right]
\]
\[
+\frac{2\pi}{\hbar{\cal V}}\,\sum_{{\bf q},\lambda}\,
\left|C_{q\lambda}\right|^2\,\left\{f_{{\bf k}-{\bf q}}\,
N_0(\omega_{q\lambda})\,
\delta(\varepsilon_{\bf k}-\varepsilon_{{\bf k}-{\bf q}}-\hbar\omega_{q\lambda})\right.
\]
\[
\left.+f_{{\bf k}+{\bf q}}\,
\left[N_0(\omega_{q\lambda})+1\right]\,
\delta(\varepsilon_{\bf k} -\varepsilon_{{\bf k}+{\bf q}}+\hbar\omega_{q\lambda})\right\}
\]
\[
+\frac{2\pi}{\hbar{\cal V}}\,\sum_{{\bf q}}\,\left|C_q\right|^2\
\left\{f_{{\bf k}-{\bf q}}\,N_0(\omega_{\rm LO})\,
\delta(\varepsilon_{\bf k}-\varepsilon_{{\bf k}-{\bf q}}-\hbar\omega_{LO})\right.
\]
\[
\left.+f_{{\bf k}+{\bf q}}\,\left[N_0(\omega_{\rm LO})+1\right]\,\delta(\varepsilon_{\bf k}
-\varepsilon_{{\bf k}+{\bf q}}+\hbar\omega_{\rm LO})\right\}
\]
\begin{equation}
+\frac{2\pi}{\hbar{\cal V}^2}\,\sum_{{\bf k}',{\bf q}}\,
\left|V_c(q)\right|^2\,
(1-f_{{\bf k}'})\,f_{{\bf k}-{\bf q}}\,f_{{\bf k}'+{\bf q}}\,
\delta(\varepsilon_{\bf k}+\varepsilon_{{\bf k}'}-\varepsilon_{{\bf k}-{\bf q}}-\varepsilon_{{\bf k}'+{\bf q}})\ ,
\label{a-2}
\end{equation}

and the scattering-out rate for electrons in the initial ${\mbox{\boldmath$k$}}$-state is

\[
{\cal W}_{\rm out}(k)=n_i\,\frac{2\pi}{\hbar{\cal V}}\,\sum_{{\bf q}}\,\left|U_i(q)\right|^2\,
\left[(1-f_{{\bf k}+{\bf q}})\,\delta(\varepsilon_{{\bf k}+{\bf q}}-\varepsilon_{\bf k})+(1-f_{{\bf k}-{\bf q}})\,
\delta(\varepsilon_{{\bf k}-{\bf q}}-\varepsilon_{\bf k})\right]
\]
\[
+\frac{2\pi}{\hbar{\cal V}}\,\sum_{{\bf q},\lambda}\,
\left|C_{q\lambda}\right|^2\,\left\{(1-f_{{\bf k}+{\bf q}})\,
N_0(\omega_{q\lambda})\,
\delta(\varepsilon_{{\bf k}+{\bf q}}-\varepsilon_{\bf k}-\hbar\omega_{q\lambda})\right.
\]
\[
\left.+(1-f_{{\bf k}-{\bf q}})\,
\left[N_0(\omega_{q\lambda})+1\right]\,
\delta(\varepsilon_{{\bf k}-{\bf q}}-\varepsilon_{\bf k}+\hbar\omega_{q\lambda})\right\}
\]
\[
+\frac{2\pi}{\hbar{\cal V}}\,\sum_{{\bf q}}\,\left|C_q\right|^2\
\left\{(1-f_{{\bf k}+{\bf q}})\,N_0(\omega_{\rm LO})\,
\delta(\varepsilon_{{\bf k}+{\bf q}}-\varepsilon_{\bf k}-\hbar\omega_{\rm LO})\right.
\]
\[
\left.+(1-f_{{\bf k}-{\bf q}})\,\left[N_0(\omega_{\rm LO})+1\right]\,\delta(\varepsilon_{{\bf k}-{\bf q}}
-\varepsilon_{\bf k}+\hbar\omega_{\rm LO})\right\}
\]
\begin{equation}
+\frac{2\pi}{\hbar{\cal V}^2}\,\sum_{{\bf k}',{\bf q}}\,
\left|V_c(q)\right|^2\,
f_{{\bf k}'}\,(1-f_{{\bf k}-{\bf q}})\,(1-f_{{\bf k}'+{\bf q}})\,
\delta(\varepsilon_{{\bf k}-{\bf q}}+\varepsilon_{{\bf k}'+{\bf q}}-\varepsilon_{\bf k}-\varepsilon_{{\bf k}'})\ .
\label{a-3}
\end{equation}
Here, $n_i$ is the volume density of ionized impurities.
For simplicity, we have introduced the notations, $f_{\bf k}\equiv f_0[\varepsilon_c({\mbox{\boldmath$k$}}),T;\,u_c]$, $\varepsilon_{\bf k}\equiv\varepsilon_c({\mbox{\boldmath$k$}})$, and $N_0(x)=[\exp(\hbar x/k_BT)-1]^{-1}$ is the Bose function for
thermal-equilibrium phonons, and $\hbar\omega_{q\lambda}$ ($\hbar\omega_{\rm LO}$) is the energy of acoustic (longitudinal-optical) phonons, respectively.
\medskip

For the electron-impurity scattering, $N_i=n_i{\cal V}$ represents the total number of impurities in the system, and

\begin{equation}
\left|U_i(q)\right|=\frac{Z^\ast e^2}{\epsilon_0\epsilon_b(q^2+Q_c^2)}\ ,
\label{a-4}
\end{equation}
where $Z^\ast$ is the charge number of fully-ionized impurity atoms.
\medskip

For the scattering of electrons with acoustic phonons, we have

\begin{equation}
\left|C_{q\ell}\right|^2=\frac{\hbar}{2\rho_i\omega_{q\ell}}\left[D_0^2q^2+\frac{9}{32}\left(eh_{14}\right)^2\right]\frac{q^2}{(q^2+Q^2_c)}\ ,
\label{a-5}
\end{equation}

\begin{equation}
\left|C_{qt}\right|^2=\frac{\hbar}{2\rho_i\omega_{qt}}\,\frac{13}{64}\left(eh_{14}\right)^2\frac{q^2}{(q^2+Q^2_c)}\ ,
\label{a-6}
\end{equation}
where $\lambda=\ell,\,t$ represents the longitudinal ($\ell$) and transverse ($t$) acoustic phonons, respectively, $\rho_i$ is the ion mass density, $D_0$ is the
deformation potential, and $h_{14}$ is the piezoelectric constant.
\medskip

For the scattering of electrons with longitudinal-optical phonons, on the other hand, we find

\begin{equation}
\left|C_q\right|^2=\frac{\hbar\omega_{\rm LO}}{2}\,\left(\frac{1}{\epsilon_{\infty}}-\frac{1}{\epsilon_s}\right)\,\frac{e^2}{\epsilon_0(q^2+Q_c^2)}\ ,
\label{a-7}
\end{equation}
where $\epsilon_{\rm s}$ and $\epsilon_{\infty}$ are the static and
high-frequency dielectric constants of the host semiconductors.
\medskip

Finally, for the scattering between two electrons, we require

\begin{equation}
V_c(q)=\frac{e^2}{\epsilon_0\epsilon_b(q^2+Q_c^2)}\ .
\label{a-8}
\end{equation}
\medskip

For the surface case, the wave vector ${\mbox{\boldmath$k$}}$ should be replaced by ${\mbox{\boldmath$k$}}_\|$, and the Coulomb potential $1/[(q^2+Q_c^2)]$ should be replaced by $\exp(-q_\|\delta_s)/2[(q_\|+q_c)]$,
where $\delta_s$ represents the thickness of the surface layer.

\section{Inverse Momentum-Relaxation-Time Tensor}
\label{app-2}

The inverse momentum-relaxation-time tensor $\tensor{\mbox{\boldmath${\tau}$}}_p^{-1}$ initially introduced in Eq.\,(\ref{dan-51}) comes from the statistically-averaged frictional forces ${\mbox{\boldmath$F$}}_x={\mbox{\boldmath$F$}}^{i}_x+{\mbox{\boldmath$F$}}^{ph}_x$
due to scattering of electrons with impurities ($i$) and phonons ($ph$).
\medskip

For electrons moving with a drift velocity $\mbox{\boldmath$v$}_d$, the frictional force ${\mbox{\boldmath$F$}}^{i}_x$ from the impurity scattering is calculated as\,\cite{r2}

\begin{equation}
{\mbox{\boldmath$F$}}^{i}_x=-n_i\,\frac{2\pi}{\hbar {\cal V}}\,\sum_{{\bf k},{\bf q}}\,\hbar{\mbox{\boldmath$q$}}\left(\hbar{\mbox{\boldmath$q$}}\cdot\mbox{\boldmath$v$}_d\right)\,\left|U_i(q)\right|^2\,
\left(-\frac{\partial f_{\bf k}}{\partial\varepsilon_{\bf k}}\right)\,\delta(\varepsilon_{{\bf k}+{\bf q}}-\varepsilon_{\bf k})\ ,
\label{b-1}
\end{equation}
and we have $\tensor{\mbox{\boldmath${\tau}$}}_{i}^{-1}\cdot\mbox{\boldmath$v$}_d=-(2/N_e)\,\tensor{\mbox{\boldmath${\cal M}$}}^{-1}\cdot{\mbox{\boldmath$F$}}^{i}_x$, where

\begin{equation}
\tensor{\mbox{\boldmath${\tau}$}}_{i}^{-1}=\frac{4\pi\hbar n_i}{\rho_0{\cal V}^2}\,\sum_{{\bf k},{\bf q}}\,\left|U_i(q)\right|^2\,
\left(-\frac{\partial f_{\bf k}}{\partial\varepsilon_{\bf k}}\right)\,\delta(\varepsilon_{{\bf k}+{\bf q}}-\varepsilon_{\bf k})\,\left\{\tensor{\mbox{\boldmath${\cal M}$}}^{-1}\otimes\left[{\mbox{\boldmath$q$}}\otimes{\mbox{\boldmath$q$}}^T\right]\right\}\ ,
\label{b-2}
\end{equation}
$\rho_0$ and $n_i$ are the volume densities of electrons and impurities, and

\[
\left[{\mbox{\boldmath$q$}}\otimes{\mbox{\boldmath$q$}}^T\right]\equiv\left[\begin{array}{ccc}
q_x^2 & q_xq_y & q_xq_z\\
q_yq_x & q_y^2 & q_yq_z\\
q_zq_x & q_zq_y & q_z^2
\end{array}\right]\ .
\]
\medskip

Physically, we can rewrite Eq.\,(\ref{b-1}) as

\begin{equation}
{\mbox{\boldmath$F$}}^{i}_x=n_i\,\sum_{{\bf q}}\,{\mbox{\boldmath$q$}}\left|U_i(q)\right|^2{\rm Im}\left[\Pi(\mbox{\boldmath$q$},\,\mbox{\boldmath$q$}\cdot\mbox{\boldmath$v$}_d)\right]\ ,
\label{new-1}
\end{equation}
where

\begin{equation}
{\rm Im}\left[\Pi(\mbox{\boldmath$q$},\,\omega)\right]=\frac{{\rm Im}\left[\Pi^{(0)}(\mbox{\boldmath$q$},\,\omega)\right]}{\left\{1-v_c(q)\,{\rm Re}\left[\Pi^{(0)}(\mbox{\boldmath$q$},\,\omega)\right]\right\}^2+\left\{v_c(q)\,{\rm Im}\left[\Pi^{(0)}(\mbox{\boldmath$q$},\,\omega)\right]\right\}^2}\ ,
\label{new-2}
\end{equation}
and $v_c(q)=e^2/\epsilon_0\epsilon_bq^2$ is the Fourier transform of a bare Coulomb potential.
Moreover, the bare polarization function $\Pi^{(0)}(\mbox{\boldmath$q$},\,\omega)$ introduced in Eq.\,(\ref{new-2}) is calculated within the random-phase approximation as

\begin{equation}
\Pi^{(0)}(\mbox{\boldmath$q$},\,\omega)=\frac{2}{{\cal V}}\,\sum_{{\bf k}}\,\frac{f_{\bf k}-f_{{\bf k}+{\bf q}}}{\hbar\omega+i0^+-\varepsilon_{{\bf k}+{\bf q}}+\varepsilon_{\bf k}}\ .
\label{new-3}
\end{equation}
Therefore, by using Eq.\,(\ref{new-1}), the inverse momentum-relaxation-time tensor $\tensor{\mbox{\boldmath${\tau}$}}_{i}^{-1}$ given by Eq.\,(\ref{b-2}) can be rewritten into the form

\begin{equation}
\tensor{\mbox{\boldmath${\tau}$}}_{i}^{-1}=-\frac{2n_i}{\rho_0{\cal V}}\,\sum_{{\bf q}}\,\left|U_i(q)\right|^2\left\{\frac{\partial{\rm Im}\left[\Pi(\mbox{\boldmath$q$},\omega)\right]}
{\partial\omega}\right\}_{\omega=\Gamma_0}
\left\{\tensor{\mbox{\boldmath${\cal M}$}}^{-1}\otimes\left[{\mbox{\boldmath$q$}}\otimes{\mbox{\boldmath$q$}}^T\right]\right\}\ ,
\label{new-4}
\end{equation}
where $\Gamma_0$ is the inverse of particle lifetime.
\medskip

Similarly, for electrons moving with a drift velocity ${\mbox{\boldmath$v$}}_d$, the frictional force ${\mbox{\boldmath$F$}}^{ph}_x$ from the acoustic and optical phonon scattering is found to be

\[
{\mbox{\boldmath$F$}}^{ph}_x=-\frac{1}{{\cal V}}\sum_{{\bf k},{\bf q},\lambda}\,\hbar{\mbox{\boldmath$q$}}\left\{\Theta^{em}_{{\bf q}\lambda}({\mbox{\boldmath$k$}})\,\left[N_0(\omega_{q\lambda})+1\right]-\Theta^{abs}_{{\bf q}\lambda}({\mbox{\boldmath$k$}})\,N_0(\omega_{q\lambda})\right\}
\]
\begin{equation}
-\frac{1}{{\cal V}}\sum_{{\bf k},{\bf q}}\,\hbar{\mbox{\boldmath$q$}}\left\{\Theta^{em}_{\bf q}({\mbox{\boldmath$k$}})\,\left[N_0(\omega_{\rm LO})+1\right]-\Theta^{abs}_{\bf q}({\mbox{\boldmath$k$}})\,N_0(\omega_{\rm LO})\right\}\ ,
\label{b-3}
\end{equation}
where the emission and absorption rates for acoustic phonons are

\begin{equation}
\Theta^{em}_{{\bf q}\lambda}({\mbox{\boldmath$k$}})=\frac{4\pi}{\hbar}\,
\left|C_{q\lambda}\right|^2(\hbar{\mbox{\boldmath$q$}}\cdot\mbox{\boldmath$v$}_d)
\left(-\frac{\partial f_{\bf k}}{\partial\varepsilon_{\bf k}}\right)\,
\delta(\varepsilon_{\bf k}-\varepsilon_{{\bf k}+{\bf q}}+\hbar\omega_{q\lambda})\ ,
\label{b-4}
\end{equation}

\begin{equation}
\Theta^{abs}_{{\bf q}\lambda}({\mbox{\boldmath$k$}})=-\frac{4\pi}{\hbar}\,
\left|C_{q\lambda}\right|^2(\hbar{\mbox{\boldmath$q$}}\cdot\mbox{\boldmath$v$}_d)
\left(-\frac{\partial f_{\bf k}}{\partial\varepsilon_{\bf k}}\right)\,
\delta(\varepsilon_{\bf k}-\varepsilon_{{\bf k}-{\bf q}}-\hbar\omega_{q\lambda})\ .
\label{b-5}
\end{equation}
In a similar way, the emission and absorption rates for longitudinal-optical phonons are calculated as

\begin{equation}
\Theta^{em}_{{\bf q}}({\mbox{\boldmath$k$}})=\frac{4\pi}{\hbar}\,
\left|C_q\right|^2(\hbar{\mbox{\boldmath$q$}}\cdot\mbox{\boldmath$v$}_d)
\left(-\frac{\partial f_{\bf k}}{\partial\varepsilon_{\bf k}}\right)\,
\delta(\varepsilon_{\bf k}-\varepsilon_{{\bf k}+{\bf q}}+\hbar\omega_{\rm LO})\ ,
\label{b-4}
\end{equation}

\begin{equation}
\Theta^{abs}_{{\bf q}}({\mbox{\boldmath$k$}})=-\frac{4\pi}{\hbar}\,
\left|C_q\right|^2(\hbar{\mbox{\boldmath$q$}}\cdot\mbox{\boldmath$v$}_d)
\left(-\frac{\partial f_{\bf k}}{\partial\varepsilon_{\bf k}}\right)\,
\delta(\varepsilon_{\bf k}-\varepsilon_{{\bf k}-{\bf q}}-\hbar\omega_{\rm LO})\ .
\label{b-5}
\end{equation}
Therefore, from Eq.\,(\ref{b-3}) we get $\tensor{\mbox{\boldmath${\tau}$}}_{ph}^{-1}\cdot\mbox{\boldmath$v$}_d=-(2/N_e)\,\tensor{\mbox{\boldmath${\cal M}$}}^{-1}\cdot{\mbox{\boldmath$F$}}^{ph}_x$, where

\[
\tensor{\mbox{\boldmath${\tau}$}}_{ph}^{-1}=\frac{8\pi\hbar}{\rho_0{\cal V}^2}\,\sum_{{\bf k},{\bf q},\lambda}\,\left|C_{q\lambda}\right|^2\,\left(-\frac{\partial f_{\bf k}}{\partial\varepsilon_{\bf k}}\right)\,
\left\{\tensor{\mbox{\boldmath${\cal M}$}}^{-1}\otimes\left[{\mbox{\boldmath$q$}}\otimes{\mbox{\boldmath$q$}}^T\right]\right\}
\]
\[
\times\left\{\left[N_0(\omega_{q\lambda})+1\right]\,\delta(\varepsilon_{\bf k}-\varepsilon_{{\bf k}+{\bf q}}+\hbar\omega_{q\lambda})
+N_0(\omega_{q\lambda})\,\delta(\varepsilon_{\bf k}-\varepsilon_{{\bf k}-{\bf q}}-\hbar\omega_{q\lambda})\right\}
\]
\[
+\frac{8\pi\hbar}{\rho_0{\cal V}^2}\,\sum_{{\bf k},{\bf q}}\,\left|C_q\right|^2\,\left(-\frac{\partial f_{\bf k}}{\partial\varepsilon_{\bf k}}\right)\,
\left\{\tensor{\mbox{\boldmath${\cal M}$}}^{-1}\otimes\left[{\mbox{\boldmath$q$}}\otimes{\mbox{\boldmath$q$}}^T\right]\right\}
\]
\begin{equation}
\times\left\{\left[N_0(\omega_{\rm LO})+1\right]\,\delta(\varepsilon_{\bf k}-\varepsilon_{{\bf k}+{\bf q}}+\hbar\omega_{\rm LO})
+N_0(\omega_{\rm LO})\,\delta(\varepsilon_{\bf k}-\varepsilon_{{\bf k}-{\bf q}}-\hbar\omega_{\rm LO})\right\}\ .
\label{b-6}
\end{equation}
\medskip

Again, we can rewrite Eq.\,(\ref{b-3}) as

\[
{\mbox{\boldmath$F$}}^{ph}_x=2\,\sum_{{\bf q},\lambda}\,{\mbox{\boldmath$q$}}\left|C_{q\lambda}\right|^2\,{\rm Im}\left[\Pi(\mbox{\boldmath$q$},\,\omega_{q\lambda}+\mbox{\boldmath$q$}\cdot\mbox{\boldmath$v$}_d)\right]
\left[N_0(\omega_{q\lambda},T)-N_0(\omega_{q\lambda}+\mbox{\boldmath$q$}\cdot\mbox{\boldmath$v$}_d,T_e)\right]
\]
\begin{equation}
+2\,\sum_{{\bf q}}\,{\mbox{\boldmath$q$}}\left|C_q\right|^2\,{\rm Im}\left[\Pi(\mbox{\boldmath$q$},\,\omega_{\rm LO}+\mbox{\boldmath$q$}\cdot\mbox{\boldmath$v$}_d)\right]
\left[N_0(\omega_{\rm LO},T)-N_0(\omega_{\rm LO}+\mbox{\boldmath$q$}\cdot\mbox{\boldmath$v$}_d,T_e)\right]\ ,
\label{new-5}
\end{equation}
and thus Eq.\,(\ref{b-6}) becomes

\[
\tensor{\mbox{\boldmath${\tau}$}}_{ph}^{-1}=-\frac{4}{\rho_0{\cal V}}\,\sum_{{\bf q},\lambda}\,\left|C_{q\lambda}\right|^2\,\left\{\frac{\partial{\rm Im}\left[\Pi(\mbox{\boldmath$q$},\omega)\right]}{\partial\omega}\right\}_{\omega=\omega_{q\lambda}}
\]
\[
\times\left[N_0(\omega_{q\lambda},T)-N_0(\omega_{q\lambda}+\mbox{\boldmath$q$}\cdot\mbox{\boldmath$v$}_d,T_e)\right]\,\left\{\tensor{\mbox{\boldmath${\cal M}$}}^{-1}\otimes\left[{\mbox{\boldmath$q$}}\otimes{\mbox{\boldmath$q$}}^T\right]\right\}
\]
\[
-\frac{4}{\rho_0{\cal V}}\,\sum_{{\bf q}}\,\left|C_q\right|^2\,\left\{\frac{\partial{\rm Im}\left[\Pi(\mbox{\boldmath$q$},\omega)\right]}{\partial\omega}\right\}_{\omega=\omega_{LO}}
\]
\begin{equation}
\times\left[N_0(\omega_{\rm LO},T)-N_0(\omega_{\rm LO}+\mbox{\boldmath$q$}\cdot\mbox{\boldmath$v$}_d,T_e)\right]\,\left\{\tensor{\mbox{\boldmath${\cal M}$}}^{-1}\otimes\left[{\mbox{\boldmath$q$}}\otimes{\mbox{\boldmath$q$}}^T\right]\right\}\ ,
\label{new-6}
\end{equation}
where $T_e$ is the temperature of hot electrons, determined from the energy-conservation equation\,\cite{r3}:

\[
\left({\mbox{\boldmath$F$}}^{i}_x+{\mbox{\boldmath$F$}}^{ph}_x\right)\cdot{\mbox{\boldmath$v$}}_d+2\,\sum_{{\bf q},\lambda}\,\left|C_{q\lambda}\right|^2\omega_{q\lambda}\,{\rm Im}\left[\Pi(\mbox{\boldmath$q$},\,\omega_{q\lambda}+\mbox{\boldmath$q$}\cdot\mbox{\boldmath$v$}_d)\right]
\left[N_0(\omega_{q\lambda},T)-N_0(\omega_{q\lambda}+\mbox{\boldmath$q$}\cdot\mbox{\boldmath$v$}_d,T_e)\right]
\]
\[
+2\,\sum_{{\bf q}}\,\left|C_q\right|^2\omega_{LO}\,{\rm Im}\left[\Pi(\mbox{\boldmath$q$},\,\omega_{LO}+\mbox{\boldmath$q$}\cdot\mbox{\boldmath$v$}_d)\right]
\left[N_0(\omega_{LO},T)-N_0(\omega_{LO}+\mbox{\boldmath$q$}\cdot\mbox{\boldmath$v$}_d,T_e)\right]=0\ .
\]

Finally, the inverse momentum-relaxation-time tensor is simply given by $\tensor{\mbox{\boldmath${\tau}$}}_{p}^{-1}=\tensor{\mbox{\boldmath${\tau}$}}_{i}^{-1}+\tensor{\mbox{\boldmath${\tau}$}}_{ph}^{-1}$.
\medskip

For the surface case, the wave vector ${\mbox{\boldmath$q$}}$ should be replaced by ${\mbox{\boldmath$q$}}_\|$, $q_z=0$ and both $\tensor{\mbox{\boldmath${\cal M}$}}_s^{-1}$ and $\tensor{\mbox{\boldmath${\tau}$}}_{sp}^{-1}$ reduce to $2\times 2$ tensors.

\section{Mobility Tensor}
\label{app-3}

From the force-balance equation in Eq.\,(\ref{dan-51}), by using the approximation $\tensor{\mbox{\boldmath${\tau}$}_{p}}^{-1}\approx(1/\tau_j)\,\delta_{ij}$ for simplicity,
we get the following group of linear equations for $\mbox{\boldmath$v$}_d=\{v_1,\,v_2,\,v_3\}$

\[
\left[1+q\tau_1\left(r_{12}B_3-r_{13}B_2\right)\right]v_1+q\tau_1\left(r_{13}B_1-r_{11}B_3\right)v_2+q\tau_1\left(r_{11}B_2-r_{12}B_1\right)v_3
\]
\begin{equation}
=q\tau_1\left(r_{11}E_1+r_{12}E_2+r_{13}E_3\right)\ ,
\label{dan-58}
\end{equation}

\[
q\tau_2\left(r_{22}B_3-r_{23}B_2\right)v_1+\left[1+q\tau_2\left(r_{23}B_1-r_{21}B_3\right)\right]v_2+q\tau_2\left(r_{21}B_2-r_{22}B_1\right)v_3
\]
\begin{equation}
=q\tau_2\left(r_{21}E_1+r_{22}E_2+r_{23}E_3\right)\ ,
\label{dan-59}
\end{equation}

\[
q\tau_3\left(r_{32}B_3-r_{33}B_2\right)v_1+q\tau_3\left(r_{33}B_1-r_{31}B_3\right)v_2+\left[1+q\tau_3\left(r_{31}B_2-r_{32}B_1\right)\right]v_3
\]
\begin{equation}
=q\tau_3\left(r_{31}E_1+r_{32}E_2+r_{33}E_3\right)\ ,
\label{dan-60}
\end{equation}
where we have used the notations ${\mbox{\boldmath$B$}}=\{B_1,\,B_2,\,B_3\}$, ${\mbox{\boldmath$E$}}=\{E_1,\,E_2,\,E_3\}$, $q=-e$ and $\tensor{\mbox{\boldmath${\cal M}$}}^{-1}=\{r_{ij}\}$.
By defining the coefficient matrix $\tensor{\mbox{\boldmath${\cal C}$}}$ for the above linear equations, i.e.,

\begin{equation}
\tensor{\mbox{\boldmath${\cal C}$}}=
\left[\begin{array}{ccc}
1+q\tau_1(r_{12}B_3-r_{13}B_2) & q\tau_1(r_{13}B_1-r_{11}B_3) & q\tau_1(r_{11}B_2-r_{12}B_1)\\
q\tau_2(r_{22}B_3-r_{23}B_2) & 1+q\tau_2(r_{23}B_1-r_{21}B_3) & q\tau_2(r_{21}B_2-r_{22}B_1)\\
q\tau_3(r_{32}B_3-r_{33}B_2) & q\tau_3(r_{33}B_1-r_{31}B_3) & 1+q\tau_3(r_{31}B_2-r_{32}B_1)
\end{array}\right]\ ,
\label{dan-61}
\end{equation}
as well as the source vector ${\mbox{\boldmath$s$}}$, given by

\begin{equation}
{\mbox{\boldmath$s$}}=\left[\begin{array}{c}
q\tau_1(r_{11}E_1+r_{12}E_2+r_{13}E_3)\\
q\tau_2(r_{21}E_1+r_{22}E_2+r_{23}E_3)\\
q\tau_3(r_{31}E_1+r_{32}E_2+r_{33}E_3)
\end{array}\right]\ ,
\label{dan-62}
\end{equation}
we can reduce the linear equations to a matrix equation $\tensor{\mbox{\boldmath${\cal C}$}}\cdot{\mbox{\boldmath$v$}}_d={\mbox{\boldmath$s$}}$ with the formal solution
$\mbox{\boldmath$v$}_d=\tensor{\mbox{\boldmath${\cal C}$}}^{-1}\cdot{\mbox{\boldmath$s$}}$. Explicitly, we find the solution $\mbox{\boldmath$v$}_d=\{v_1,\,v_2,\,v_3\}$
for $j=1,\,2,\,3$ by

\begin{equation}
v_j=\frac{Det\{\tensor{\mbox{\boldmath${\Delta}$}}_j\}}{Det\{\tensor{\mbox{\boldmath${\cal C}$}}\}}\ ,
\label{dan-63}
\end{equation}
where

\footnotesize
\begin{equation}
\tensor{\mbox{\boldmath${\Delta}$}}_1=
\left[\begin{array}{ccc}
q\tau_1(r_{11}E_1+r_{12}E_2+r_{13}E_3) & q\tau_1(r_{13}B_1-r_{11}B_3) & q\tau_1(r_{11}B_2-r_{12}B_1)\\
q\tau_2(r_{21}E_1+r_{22}E_2+r_{23}E_3) & 1+q\tau_2(r_{23}B_1-r_{21}B_3) & q\tau_2(r_{21}B_2-r_{22}B_1)\\
q\tau_3(r_{31}E_1+r_{32}E_2+r_{33}E_3) & q\tau_3(r_{33}B_1-r_{31}B_3) & 1+q\tau_3(r_{31}B_2-r_{32}B_1)
\end{array}\right]\ ,
\label{dan-64}
\end{equation}

\begin{equation}
\tensor{\mbox{\boldmath${\Delta}$}}_2=
\left[\begin{array}{ccc}
1+q\tau_1(r_{12}B_3-r_{13}B_2) & q\tau_1(r_{11}E_1+r_{12}E_2+r_{13}E_3) & q\tau_1(r_{11}B_2-r_{12}B_1)\\
q\tau_2(r_{22}B_3-r_{23}B_2) & q\tau_2(r_{21}E_1+r_{22}E_2+r_{23}E_3) & q\tau_2(r_{21}B_2-r_{22}B_1)\\
q\tau_3(r_{32}B_3-r_{33}B_2) & q\tau_3(r_{31}E_1+r_{32}E_2+r_{33}E_3) & 1+q\tau_3(r_{31}B_2-r_{32}B_1)
\end{array}\right]\ ,
\label{dan-65}
\end{equation}

\begin{equation}
\tensor{\mbox{\boldmath${\Delta}$}}_3=
\left[\begin{array}{ccc}
1+q\tau_1(r_{12}B_3-r_{13}B_2) & q\tau_1(r_{13}B_1-r_{11}B_3) & q\tau_1(r_{11}E_1+r_{12}E_2+r_{13}E_3)\\
q\tau_2(r_{22}B_3-r_{23}B_2) & 1+q\tau_2(r_{23}B_1-r_{21}B_3) & q\tau_2(r_{21}E_1+r_{22}E_2+r_{23}E_3)\\
q\tau_3(r_{32}B_3-r_{33}B_2) & q\tau_3(r_{33}B_1-r_{31}B_3) & q\tau_3(r_{31}E_1+r_{32}E_2+r_{33}E_3)
\end{array}\right]\ .
\label{dan-66}
\end{equation}
\medskip

\normalsize
By assuming $r_{ij}=0$ for $i\neq j$, $r_{jj}=1/m_j^\ast$ and introducing the notation $\mu_j=q\tau_j/m_j^\ast$, we find

\begin{equation}
\tensor{\mbox{\boldmath${\cal C}$}}=
\left[\begin{array}{ccc}
1 & -\mu_1B_3 & \mu_1B_2\\
\mu_2B_3 & 1 & -\mu_2B_1\\
-\mu_3B_2 & \mu_3B_1 & 1
\end{array}\right]\ ,
\label{dan-67}
\end{equation}

\begin{equation}
\tensor{\mbox{\boldmath${\Delta}$}}_1=
\left[\begin{array}{ccc}
\mu_1E_1 & -\mu_1B_3 & \mu_1B_2\\
\mu_2E_2 & 1 & -\mu_2B_1\\
\mu_3E_3 & \mu_3B_1 & 1
\end{array}\right]\ ,
\label{dan-68}
\end{equation}

\begin{equation}
\tensor{\mbox{\boldmath${\Delta}$}}_2=
\left[\begin{array}{ccc}
1 & \mu_1E_1 & \mu_1B_2\\
\mu_2B_3 & \mu_2E_2 & -\mu_2B_1\\
-\mu_3B_2 & \mu_3E_3 & 1
\end{array}\right]\ ,
\label{dan-69}
\end{equation}

\begin{equation}
\tensor{\mbox{\boldmath${\Delta}$}}_3=
\left[\begin{array}{ccc}
1 & -\mu_1B_3 & \mu_1E_1\\
\mu_2B_3 & 1 & \mu_2E_2\\
-\mu_3B_2 & \mu_3B_1 & \mu_3E_3
\end{array}\right]\ ,
\label{dan-70}
\end{equation}
and
\[
Det\{\tensor{\mbox{\boldmath${\cal C}$}}\}=1+(B_1^2\mu_2\mu_3+B_2^2\mu_3\mu_1+B_3^2\mu_1\mu_2)\ ,
\]

\[
Det\{\tensor{\mbox{\boldmath${\Delta}$}}_1\}=\mu_1E_1+\mu_1(B_3E_2\mu_2-B_2E_3\mu_3)+\mu_1\mu_2\mu_3B_1({\bf E}\cdot{\bf B})\ ,
\]

\[
Det\{\tensor{\mbox{\boldmath${\Delta}$}}_2\}=\mu_2E_2+\mu_2(B_1E_3\mu_3-B_3E_1\mu_1)+\mu_1\mu_2\mu_3B_2({\bf E}\cdot{\bf B})\ ,
\]

\[
Det\{\tensor{\mbox{\boldmath${\Delta}$}}_3\}=\mu_3E_3+\mu_3(B_2E_1\mu_1-B_1E_2\mu_2)+\mu_1\mu_2\mu_3B_3({\bf E}\cdot{\bf B})\ .
\]
For a special case with ${\mbox{\boldmath$B$}}=\{0,\,0,\,B_3\}$, we get

\begin{equation}
\tensor{\mbox{\boldmath${\mu}$}}(B_3)=\frac{1}{1+\mu_1\mu_2B_3^2}\,
\left[\begin{array}{ccc}
\mu_1 & \mu_1\mu_2B_3 & 0\\
-\mu_2\mu_1B_3 & \mu_2 & 0\\
0 & 0 & \mu_3(1+\mu_1\mu_2B_3^2)
\end{array}\right]\ .
\label{dan-72}
\end{equation}
\medskip

If we further assume $m_1^\ast=m_2^\ast=m_3^\ast=m^\ast$ and $\tau_1=\tau_2=\tau_3=\tau_p$, we obtain
$Det\{\tensor{\mbox{\boldmath${\cal C}$}}\}=1+\mu^2_0B^2$,
$Det\{\tensor{\mbox{\boldmath${\Delta}$}}_1\}=-\mu_0E_1+\mu_0^2(B_3E_2-B_2E_3)-\mu_0^3B_1({\mbox{\boldmath$E$}}\cdot{\mbox{\boldmath$B$}})$,
$Det\{\tensor{\mbox{\boldmath${\Delta}$}}_2\}=-\mu_0E_2+\mu_0^2(B_1E_3-B_3E_1)-\mu_0^3B_2({\mbox{\boldmath$E$}}\cdot{\mbox{\boldmath$B$}})$, and
$Det\{\tensor{\mbox{\boldmath${\Delta}$}}_3\}=-\mu_0E_3+\mu_0^2(B_2E_1-B_1E_2)-\mu_0^3B_3({\mbox{\boldmath$E$}}\cdot{\mbox{\boldmath$B$}})$,
where $\mu_0=e\tau_p/m^\ast$. As a result, the mobility tensor $\tensor{\mbox{\boldmath${\mu}$}}({\mbox{\boldmath$B$}})$, which is defined through $\mbox{\boldmath$v$}_d=\tensor{\mbox{\boldmath${\mu}$}}({\mbox{\boldmath$B$}})\cdot{\mbox{\boldmath$E$}}$, can be written as

\begin{equation}
\tensor{\mbox{\boldmath${\mu}$}}({\mbox{\boldmath$B$}})=-\frac{\mu_0}{1+\mu^2_0B^2}\,
\left[\begin{array}{ccc}
1+\mu_0^2B_1^2 & -\mu_0B_3+\mu_0^2B_1B_2 & \mu_0B_2+\mu_0^2B_1B_3\\
\mu_0B_3+\mu_0^2B_2B_1 & 1+\mu_0^2B_2^2 & -\mu_0B_1+\mu_0^2B_2B_3\\
-\mu_0B_2+\mu_0^2B_3B_1 & \mu_0B_1+\mu_0^2B_3B_2 & 1+\mu_0^2B_3^2
\end{array}\right]\ ,
\label{dan-71}
\end{equation}
where $B^2=B_1^2+B_2^2+B_3^2$.
\medskip

For the surface case with $E_3=0$ and $v_3=0$, $\tensor{\mbox{\boldmath${\cal M}$}}_s^{-1}$, $\tensor{\mbox{\boldmath${\tau}$}}_{sp}^{-1}$ and $\tensor{\mbox{\boldmath${\mu}$}}_s({\mbox{\boldmath$B$}})$ all reduce to $2\times 2$ tensors.

\section{Bulk and Surface Conductivity Tensors}

Under a parallel external electric field ${\mbox{\boldmath${E}$}}=(E_x,E_y,0)$ and a perpendicular magnetic field ${\mbox{\boldmath${B}$}}=(0,0,B)$,
the total parallel current per length in a $p$-$n$ junction structure is given by
$\displaystyle{\int_{-L_A}^{L_D} dz\,\left[{\mbox{\boldmath${j}$}}^\|_c(z)+{\mbox{\boldmath${j}$}}^\|_v(z)\right]+{\mbox{\boldmath${j}$}}_s^{\pm}}$, where $L_D$ and $L_A$ are the distribution ranges for donors and acceptors, respectively.
Here, by using the second-order Boltzmann moment equation, the bulk current densities are found to be\,\cite{r2}

\begin{equation}
{\mbox{\boldmath${j}$}}_{c,v}^\|(z)=\frac{2e\gamma_{e,h}m_{e,h}^\ast\tau_{e,h}(z)}{\tau_{p(e,h)}(z)}\,{\mbox{\boldmath${v}$}}^\|_{c,v}[u_{c,v}(z)]
\left\{\left[\tensor{\mbox{\boldmath${\mu}$}}^\|_{c,v}({\mbox{\boldmath${B}$}},z)\cdot{\mbox{\boldmath${E}$}}\right]\right\}\cdot{\mbox{\boldmath${v}$}}^\|_{c,v}[u_{c,v}(z)]\,
{\cal D}_{c,v}[u_{c,v}(z)]\ ,
\label{app-19}
\end{equation}
where ${\cal D}_{c,v}[u_{c,v}(z)]=(\sqrt{u_{c,v}(z)}/4\pi^2)\,(2m_{e,h}^\ast/\hbar^2)^{3/2}$ is the electron and hole density-of-states per spin, $u_{c,v}(z)=(\hbar k_F^{e,h})^2/2m^*_{e,h}$
and $k_F^{e,h}$ are Fermi energies and wave vectors in a bulk,
$m^*_{e,h}$ are effective masses of electrons and holes,
$\tau_{e,h}(z)$ and $\tau_{p(e,h)}(z)$ are bulk energy- and momentum-relaxation times,
${\mbox{\boldmath${v}$}}^\|_{c,v}(\mbox{\boldmath${k}$})=-\gamma_{e,h}\,\hbar{\mbox{\boldmath${k}$}}_\|/m^\ast_{e,h}$,
and $\gamma_{e,h}=-1$ (electrons) and $+1$ (holes), respectively. Similarly, the surface current per length is

\begin{equation}
{\mbox{\boldmath${j}$}}^{\pm}_s=\mp\color{black}{\frac{e\tau_sm_s^*}{\tau_{sp}}}\color{black}\,{\mbox{\boldmath${v}$}}^{\pm}_s(u_s)
\left\{\left[\tensor{\mbox{\boldmath${\mu}$}}^{\pm}_s({\mbox{\boldmath${B}$}})\cdot{\mbox{\boldmath${E}$}}\right]\right\}\cdot{\mbox{\boldmath${v}$}}^{\pm}_s(u_s)\,
\rho_s(u_s)\ ,
\label{app-20}
\end{equation}
where \color{black}{$\rho_s(u_s)=\Delta_0/(2\pi\hbar^2v_F^2)$ and $u_s=(\hbar k_F^sv_F)^2/2\Delta_0$} \color{black}are the surface density-of-states and Fermi energy, $k^s_F=\sqrt{4\pi \sigma_s}$,
$v_F$ is the Fermi velocity of a Dirac cone,
$\tau_{s}$ and $\tau_{sp}$ are surface energy- and momentum relaxation times,
and \color{black}{${\mbox{\boldmath${v}$}}^{\pm}_s(\mbox{\boldmath${k}$}_\|)=\pm\hbar v^2_F{\mbox{\boldmath${k}$}}_\|/\Delta_0$.}
\medskip

\color{black}
From Eq.\,(\ref{app-19}), we find the bulk conductivity tensor as

\begin{equation}
\tensor{\mbox{\boldmath${\sigma}$}}^\|_{c,v}({\mbox{\boldmath${B}$}})
=e\gamma_{e,h}\,\int_{-L_A}^{L_D} dz\,n_{e,h}(z)\,\left[\frac{\tau_{e,h}(z)}{\tau_{p(e,h)}(z)}\right]\,\tensor{\mbox{\boldmath${\mu}$}}^\|_{c,v}({\mbox{\boldmath${B}$}},z)\ .
\label{app-21}
\end{equation}
On the other hand, from Eq.\,(\ref{app-20}) we get the surface conductivity tensor, given by

\begin{equation}
\tensor{\mbox{\boldmath${\sigma}$}}^{\pm}_{s}({\mbox{\boldmath${B}$}})=e\sigma_s\left(\frac{\tau_s}{\tau_{sp}}\right)\tensor{\mbox{\boldmath${\mu}$}}^{\pm}_s({\mbox{\boldmath${B}$}})\ .
\label{app-22}
\end{equation}
Therefore, the total conductivity tensor $\tensor{\mbox{\boldmath${\sigma}$}}_{tot}({\mbox{\boldmath${B}$}})=\tensor{\mbox{\boldmath${\sigma}$}}^\|_{c}({\mbox{\boldmath${B}$}})+\tensor{\mbox{\boldmath${\sigma}$}}^\|_{v}({\mbox{\boldmath${B}$}})
+\tensor{\mbox{\boldmath${\sigma}$}}^{\pm}_{s}({\bf B})$ can be obtained from

\[
\tensor{\mbox{\boldmath${\sigma}$}}_{tot}({\mbox{\boldmath${B}$}})=e\,\tensor{\mbox{\boldmath${\mu}$}}^\|_{v}({\mbox{\boldmath${B}$}})N_AA_h\left[(L_A-W_p)+\int^{W_p}_{0} dz\,\exp\left(-\frac{\beta e\bar{\mu}_hN_A}{2\epsilon_0\epsilon_rD_h}\,z^2\right)\right]
-e\,\tensor{\mbox{\boldmath${\mu}$}}^\|_{c}({\mbox{\boldmath${B}$}})N_DA_e
\]
\begin{equation}
\times\left[(L_D-W_n)+\int^{W_n}_{0} dz\,\exp\left(-\frac{\beta e\bar{\mu}_eN_D}{2\epsilon_0\epsilon_rD_e}\,z^2\right)\right]+\color{black}{e\,\tensor{\mbox{\boldmath${\mu}$}}^{\pm}_s({\mbox{\boldmath${B}$}})\,\left(\frac{\alpha_0\Delta_0}{2\pi\hbar^2v_F^2}\right)\left(L_A-L_0\right)A_s\ ,}
\label{app-23}
\end{equation}
\color{black}
where $\alpha_0$ and $L_0$ are constants to be determined experimentally, $N_{D,A}$ are doping concentrations,
$W_n$ and $W_p$ are depletion ranges for donors and acceptors in a $p$-$n$ junction,
$\bar{\mu}_{e,h}$ are $\mu_0(z)$ evaluated at $n_{e,h}(z)=N_{D,A}$, $D_{e,h}$ are diffusion coefficients,
and $\beta=4/3$ ($\beta=7/3$) for longitudinal (Hall) conductivity. In addition,
the averaged mobilities $\tensor{\mbox{\boldmath${\mu}$}}^\|_{c,v}({\mbox{\boldmath${B}$}})$ are defined by
their values of $\tau_{p(e,h)}(z)$ at $n_{e,h}(z)=N_{D,A}$, and three introduced coefficients are $A_s=\tau_s/\tau_{sp}\approx 3/4$,

\[
A_{e,h}=\left.\frac{\tau_{e,h}(z)}{\tau_{p(e,h)}(z)}\right|_{n_{e,h}(z)=N_{D,A}}
\]
\begin{equation}
=\frac{1}{6}\left(\frac{Q_c}{k_F^{e,h}}\right)^2\left[2\ln\left(\frac{2k_F^{e,h}}{Q_c}\right)-1\right]
=\frac{Q^2_c}{6(3\pi^2N_{D,A})^{2/3}}\left\{2\ln\left[\frac{2(3\pi^2N_{D,A})^{1/3}}{Q_c}\right]-1\right\}\ ,
\label{app-24}
\end{equation}
where $1/Q_c$ is the Thomas-Fermi screening length.
\medskip

In addition, the bulk energy-relaxation times $\tau_{e,h}(z)$ are calculated as
\[
\frac{1}{\tau_{e,h}(z)}=\left[\frac{2n_i}{n_{e,h}(z)\pi\hbar Q_c^2}\right]\left(\frac{e^2}{\epsilon_0\epsilon_r}\right)^2\int_0^{k^{e,h}_F(z)} dk\,{\cal D}_{c,v}(\varepsilon^{c,v}_k)\left(\frac{4k^2}{4k^2+Q_c^2}\right)
\]
\begin{equation}
=\left[\frac{n_im^*_{e,h}}{8n_{e,h}(z)\pi^3\hbar^3Q_c^2}\right]\left(\frac{e^2}{\epsilon_0\epsilon_r}\right)^2
\left\{[2k_F^{e,h}(z)]^2-Q_c^2\ln\left(\frac{[2k_F^{e,h}(z)]^2+Q_c^2}{Q_c^2}\right)\right\}\ ,
\label{app-25}
\end{equation}
and the surface energy-relaxation time $\tau_s$ is found to be

\begin{equation}
\frac{1}{\tau_s}=\frac{2\sigma_i}{\pi^2\sigma_s\hbar^2v_F}\left(\frac{e^2}{2\epsilon_0\epsilon_r}\right)^2
\int_0^{\pi} d\phi\,\int_0^{k_F^s}\,\frac{k^2_\|\,dk_\|}{(q_c+2k_\||\cos\phi|)^2}\ ,
\label{app-26}
\end{equation}
where $n_i$ and $\sigma_i$ are the concentration and surface density of impurities, respectively.
\medskip

Finally, the bulk chemical potentials for electrons [$u_c(z)$] and holes [$u_v(z)$] are calculated as

\begin{equation}
\left[u_{c,v}(z)\right]^{3/2}=3\pi^2\left(\frac{h^2}{2m^\ast_{e,h}}\right)^{3/2}n_{e,h}(z)\ ,
\label{app-27}
\end{equation}
and the carrier density functions are

\begin{equation}
n_{e,h}(z)=N_{D,A}\exp\left\{-\gamma_{e,h}\left(\frac{\bar{\mu}_{e,h}}{D_{e,h}}\right)\left[\Phi(z)+\gamma_{e,h}(E_F^{e,h}/e)\right]\right\}\ .
\label{app-28}
\end{equation}
Here, the expression for the introduced potential function $\Phi(z)$ is given by

\begin{equation}
\Phi(z)=\left\{\begin{array}{cc}
-E_F^h/e\ , & z<-W_p\\
-E_F^h/e+(eN_A/2\epsilon_0\epsilon_r)\,(z+W_p)^2\ , & -W_p<z<0\\
E_F^e/e-(eN_D/2\epsilon_0\epsilon_r)\,(W_n-z)^2\ , & 0<z<W_n\\
E_F^e/e\ , & z>W_n
\end{array}\right.\ ,
\label{app-29}
\end{equation}
and $E_F^e$ ($E_F^h$) is the Fermi energy of electrons (holes) at zero temperature and defined far away from the depletion region.

\end{document}